\documentclass[aps,prd,twocolumn,superscriptaddress,showpacs,showkeys,nofootinbib]{revtex4-1}

\usepackage{latexsym}
\usepackage{amsmath}
\usepackage{amssymb}
\usepackage{amsfonts}

\usepackage{color}
\usepackage{soul}

\usepackage{supertabular} 
\usepackage{placeins}
\usepackage{epsfig}
\usepackage{graphicx}

\begin{document}


\title{The $Z_c$ structures in a coupled-channels model}

\author{Pablo G. Ortega}
\email[]{pgortega@usal.es}
\affiliation{Grupo de F\'isica Nuclear and Instituto Universitario de F\'isica 
Fundamental y Matem\'aticas (IUFFyM), Universidad de Salamanca, E-37008 Salamanca, Spain}

\author{Jorge Segovia}
\email[]{jsegovia@ifae.es}
\affiliation{Grup de F\'isica Te\`orica, Dept. F\'isica and IFAE-BIST, Universitat Aut\`onoma de Barcelona, E-08193 Bellaterra (Barcelona), Spain}

\author{David R. Entem}
\email[]{entem@usal.es}
\affiliation{Grupo de F\'isica Nuclear and Instituto Universitario de F\'isica Fundamental y Matem\'aticas (IUFFyM), Universidad de Salamanca, E-37008 Salamanca, Spain}

\author{Francisco Fern\'andez}
\email[]{fdz@usal.es}
\affiliation{Grupo de F\'isica Nuclear and Instituto Universitario de F\'isica 
Fundamental y Matem\'aticas (IUFFyM), Universidad de Salamanca, E-37008 
Salamanca, Spain}


\date{\today}

\begin{abstract}
The $Z_{c}(3900)^\pm/Z_c(3885)^\pm$ and $Z_{c}(4020)^\pm$ are two charmonium-like structures discovered in the $\pi J/\psi$ and $D^\ast\bar D^{(\ast)}+h.c.$ invariant mass spectra. Their nature is puzzling due to their charge, which forces its minimal quark content to be $c\bar c u\bar d$ ($c\bar c d\bar u$). Thus, it is necessary to explore four-quark systems in order to understand their inner structure. Additionally, their strong coupling to channels such as $\pi J/\psi$ and the closeness of their mass to $D^\ast\bar D^{(\ast)}$-thresholds stimulates both a molecular interpretation or a coupled-channels threshold effect. In this work we perform a coupled-channels calculation of the $I^G(J^{PC})=1^+(1^{+-})$ sector including $D^{(\ast)}\bar D^{\ast}+h.c.$, $\pi J/\psi$ and $\rho\eta_c$ channels in the framework of a constituent quark model which satisfactorily describes a wide range of properties of (non-)conventional hadrons containing heavy quarks. The meson-meson interactions are dominated by the non-diagonal $\pi J/\psi-D^\ast\bar D^{(\ast)}$ and $\rho\eta_c-D^\ast\bar D^{(\ast)}$ couplings which indicates that the $Z_{c}(3900)^\pm/Z_c(3885)^\pm$ and $Z_{c}(4020)^\pm$ are unusual structures. The study of the analytic structure of the $S$-matrix allows us to conclude that the point-wise behavior of the line shapes in the $\pi J/\psi$ and $D\bar D^*$ invariant mass distributions is due to the presence of two virtual states that produce the $Z_c$ peaks.
\end{abstract}



\maketitle


\section{Introduction} \label{sec:introduction}

Since early 2000s, signals of non-conventional meson structures have appeared in the so-called B-factories and other accelerator facilities. In 2003, the $X(3872)$ was discovered by Belle~\cite{Choi:2003ue}, promptly confirmed by  BaBar~\cite{Aubert:2004ns}, CDF~\cite{Acosta:2003zx} and D0~\cite{Abazov:2004kp} Collaborations, whereas BaBar and CLEO Collaborations reported the observation of two puzzling heavy-light mesons: $D_{s0}^\ast(2317)$ and $D_{s1}(2460)$~\cite{Aubert:2003fg, Besson:2003cp}. The common characteristic of all these states is that, although their quantum numbers are compatible with naive $q\bar q$ structures, their masses and decay properties point to more complex structures involving higher Fock-state components.

However, it was not until 2011 when undeniable evidences of exotic mesons with forbidden quantum numbers for a quark-antiquark pair were observed. These indications were found by the Belle Collaboration~\cite{Belle:2011aa, Adachi:2012cx} in the bottom sector with the discovery of the $Z_{b}(10610)$ and $Z_{b}(10650)$ charged structures in the $\Upsilon(5S)\to \pi^+\pi^-\Upsilon(nS)$ reaction. As the reader can guess, they are close to the $B\bar B^\ast$ and $B^\ast \bar B^\ast$ thresholds, respectively. 

One of the partners of such structures in the charmonium sector arrived few years later, in 2013, when the BESIII and Belle Collaborations claimed the observation of a charged structure, dubbed $Z_{c}(3900)$ ~\cite{Ablikim:2013mio, Liu:2013dau}, in the $\pi^+\pi^- J/\psi$ invariant mass spectrum of the $e^+e^-\rightarrow\pi^+\pi^- J/\psi$ process at $\sqrt{s}=4.26$ GeV. It was further confirmed in the same channel but at $\sqrt{s}=4.17$ GeV in Ref.~\cite{Xiao:2013iha} by a reanalysis of CLEO-c data. Its neutral partner was also reported in Refs.~\cite{Xiao:2013iha, Ablikim:2015tbp}.
	
The $Z_{c}(3900)$ mass from the $\pi J/\psi$ invariant mass spectrum measurement, $M=(3899.0\pm 3.6\pm 4.9)$ MeV/$c^2$~\cite{Ablikim:2013mio}, is 24 MeV/$c^2$ above the $D\bar D^*$ mass threshold and thus it is natural to observe the $Z_{c}(3900)$ in the $(D\bar D^*)_{I=1}$ channel. This was confirmed by the BESIII Collaboration~\cite{Ablikim:2013xfr} which reported a strong threshold enhancement in the $D\bar D^*$ invariant mass distribution of the $e^+e^-\rightarrow\pi^\pm (D\bar D^*)^\mp$ process at $\sqrt{s}=4.26$ GeV, referred as $Z_c(3885)$. Fitted with a mass-dependent-width Breit-Wigner line shape, its mass and width were determined to be $M=(3883.9\pm 1.5\pm 4.2)$ MeV/$c^2$ and $\Gamma = (24.8\pm 3.3\pm 11.0)$ MeV, the data being compatible with the one expected for $J^P=1^+$ quantum numbers. A few years later, such determination was updated to $M=(3881.7\pm1.6\pm6.9)$ MeV/c$^2$ and $\Gamma=(26.6\pm2.0\pm2.1)$ MeV, with measurements at $\sqrt{s}=4.26$ GeV and $\sqrt{s}=4.23$ GeV~\cite{Ablikim:2015swa}. Recently, the spin and parity of the $Z_c(3900)$ was analyzed in Ref.~\cite{Collaboration:2017njt} with the result that the $J^P=1^+$ assignement is favoured by the data; moreover, its mass was reduced to $M=(3881.2\pm 4.2\pm 52.7)$ and the new width update to $\Gamma=(51.8\pm 4.6\pm 36.0)$ MeV, using a Flatt\'e-like expresion. This new determination supports the idea that the $Z_c(3900)$/$Z_c(3885)$ signals correspond to an unique state.

Soon after all these experimental activities, the BESIII Collaboration reported the discovery of another charged state, the $Z_{c}(4020)$ resonance, in the $e^+e^-\to \pi^+\pi^- h_c$ channel with a mass of $M=(4022.9\pm 0.8\pm 2.7)$ MeV/$c^2$ and a width of $\Gamma=(7.9\pm 2.7\pm 2.6)$ MeV~\cite{Ablikim:2013wzq}. Later on, a similar structure with a mass of $M=(4026.3\pm 2.6\pm 3.7)$ MeV/$c^2$ and a width of $\Gamma= (24.8\pm 5.6\pm 5.7)$ MeV was observed by the BESIII Collaboration in the process $e^+e^-\rightarrow\pi^\pm (D^*\bar D^*)^\mp$ at $\sqrt{s}=4.26$ GeV~\cite{Ablikim:2013emm}. The neutral partner of the $Z_c(4020)$ was reported by the BESIII Collaboration in Ref.~\cite{Ablikim:2014dxl}.

The fact that most of these structures have been found at $\sqrt{s}=4.26$ GeV led several authors to think that the $Y(4260)$ resonance may be responsible for the production of the $Z_c(3900)$ and $Z_c(4020)$ structures due to their unconventional properties. However, the detection of the $Z_c(3900)$ at the peak of the $\psi(4160)$ state by the CLEO-c Collaboration discarded any effect of the $Y(4260)$ resonance~\cite{Xiao:2013iha}.

The $Z_c(3900)$ and $Z_c(4020)$ states are charged and close to $D\bar D^*$ and $D^*\bar D^*$ thresholds, respectively. Therefore, they cannot be described as pure $q\bar q$ states and it is expected that $D^{(*)}\bar D^*$ components are significant in their wave functions. This fact can be implemented in different theoretical scenarios, ranging from hadron molecules~\cite{Guo:2013sya, Mehen:2013mva, He:2013nwa, Liu:2014eka}, to tetraquark structures~\cite{Esposito:2014rxa, Dias:2013xfa, Agaev:2017tzv, Wang:2013vex, Qiao:2013raa, Deng:2014gqa} or simple kinematic effects linked to the opening of meson-meson thresholds~\cite{Swanson:2014tra, Swanson:2015bsa}.

At this point, it is worth noticing that not all enhancements of the cross section correspond to a resonance. As discussed by R.~G.~Newton~\cite{Newton:1982qc}, the vicinity of the threshold of a new opening channel causes the cross section to show an anomalous behavior in form of a cusp. From a mathematical point of view, this behavior is produced by the branch-point of the $S$-matrix at threshold. Resonances are also enhancements of the cross section, but corresponding to poles of the $S$-matrix on the unphysical Riemann sheet close to the physical one. In this case, the cross section can be parametrized with a Breit-Wigner form with two parameters: the energy (mass) and the width. Furthermore, besides bound states which correspond to poles of the $S$-matrix on the real energy axis of the physical Riemann sheet below the lowest threshold, the $S$-matrix may present poles on the real energy axis but in the unphysical Riemann sheet. If these virtual states have sufficiently low energy, then their presence modifies appreciably the energy dependence of the scattering cross section which can be anomalously large. Therefore, to correctly interpret the point-wise behavior of the line shapes of a particular process one must look carefully on the analytical properties of the $S$-matrix.

In a couple of papers, E.~S.~Swanson~\cite{Swanson:2014tra, Swanson:2015bsa} claimed that the $Z_c$ and $Z_b$ resonances could arise from kinematic threshold effects, meaning that the $D\bar D^\ast$ interaction, in case of $Z_c(3900)$, is strong enough to cause an enhancement just above the $S$-wave threshold. The fact that the masses of such resonances are slightly above their respective thresholds seems to favor this possibility. Swanson used a simplified model to describe Belle and BESIII data as kinematic cusps, successfully capturing the features of all the data and indicating that there is no evidence for strong $D\bar D^\ast$ or $D^\ast\bar D^\ast$ rescattering in the $(I=1)$ $J^{PC}=1^{+-}$ channel. Hence, he concluded that isovector rescattering is not sufficiently attractive to generate dynamical bound states and so exotic resonances are not required to explain the data.

Kinematic cusp interpretations were subsequently challenged by F.-K.~Guo {\it et al.}~\cite{Guo:2014iya}. Using a non-relativistic effective field theory, the authors argued that a cusp always exists due to the opening of an $S$-wave threshold but, in order to reproduce the narrow and pronounced peak observed by experimentalists, a non-perturbative interaction amongst the heavy mesons is necessary and thus a nearby pole in the $S$-matrix must appear. This motivated a detailed analysis of the experimental data using reaction theory and considering different scenarios for the inner structure of the $Z_c(3900)$~\cite{Pilloni:2016obd}. The authors claimed that a cusp is not strong enough to reproduce the peak while a molecular or virtual state can describe equally well the experimental data.

Further calculations with different effective field theories have been performed. F.~Aceti {\it et al.}~\cite{Aceti:2014uea}, using the local hidden gauge approach, studied the $D\bar D^*$ interaction in a coupled channel Bethe-Salpeter calculation. They found in the $I=1$ channel a barely $D\bar D^*$ bound state decaying into $\rho\eta_c$ and $\pi J/\psi$ channels with a mass of $3869-3875$ MeV. Additionally, M.~Albaladejo {\it et al.}~\cite{Albaladejo:2015lob} performed a simultaneous study of the invariant mass distribution for the $\pi J/\psi$ and $D\bar D^*$ channels and found different interpretations when exploring with an energy dependent or independent $D\bar D^\ast$ $S$-wave interaction. In the first case, the $Z_c$ enhancement was originated from a resonance with a mass around the $D\bar D^\ast$ threshold. In the second one, the $Z_c$ signal was produced by a virtual state which must have a hadronic molecular nature. In any case, a $D\bar D^\ast$ bound state was not favored. It is worth remarking that, following these authors, the tetraquark scenario would be discarded if the $S$-matrix pole is at the unphysical sheet (virtual state), as it would originate from $D\bar D^\ast$ interaction solely. A similar conclusion can be found in Ref.~\cite{He:2017lhy},~\cite{He:2013nwa} where the authors look for the $Z_c(3900)$ and $Z_c(4020)$ state in  the $\pi J/\psi$ and $D\bar D^*$  and the $D^*\bar D^*$ invariant mass spectra respectively. The calculation was done in a quasi-potential Bethe-Salpeter equation approach at the hadronic level. They found that the $\pi J/\psi$-$D\bar D^*$ interaction produces a virtual state at a mass around $3870$ MeV whereas they found a bound state for the $Z_c(4020)$ with reasonable values of the parameters. 
Additionally, the authors of Ref.~\cite{Aceti:2014kja} study the $D^\ast\bar D^\ast$ dynamics in the local hidden gauge approach, focusing on the two pion exchange, and found that such $Z_c$ states are in the border between weakly-bound and virtual states.

Many papers have been released in the last years reporting exploratory LQCD studies of the $Z_c$ structures~\cite{Prelovsek:2014swa, Prelovsek:2013xba, Chen:2014afa, Chen:2015jwa, Lee:2014uta, Ikeda:2016zwx}. S.~Prelovsek {\it et al.}~\cite{Prelovsek:2014swa, Prelovsek:2013xba} looked for the $Z_c(3900)^+$ state in the energy region below $4$~GeV, finding only two-particle $\pi J/\psi$ and $D\bar D^\ast$ scattering states and no signal of the $Z_c(3900)^+$. In a similar search  Y.~Chen{\it} et al.~\cite{Chen:2014afa} found that the $D\bar D^\ast$ interaction is weakly repulsive and, therefore, the results do not support the possibility of a shallow bound state at least at the pion mass values studied. A different approach is taken by Y. Ikeda {\it et al.}~\cite{Ikeda:2016zwx}. They performed a coupled-channels calculation taking into account the $\pi J/\psi$, $\rho\eta_c$ and $DD^*$ channels and finding that the interactions between them are dominated by off-diagonal couplings which may indicate that the $Z_c(3900)^+$ can be explained as a threshold cusp. There are technicalities involving LQCD computations, such as large pion masses, small volumes and a set of interpolators not large enough for having overlap with the physical state, which still prevent to make a definitive statement. Nonetheless, seems that the available LQCD calculations are robust enough to discard the bound state option for the $Z_c^+$ states.

From a molecular point of view, the fact that the $Z_c$ structures have $I=1$ makes that the $D^{(\ast)}\bar D^{(\ast)}$ interaction weak because in these channels the one-pion-interaction isospin coefficient is a factor 3 smaller than in the $I=0$ sector. This leaves the coupled channel calculations as the most promising option to produce some resonance, bound or virtual state, if any.

In this work we analyze the $I^G(J^{PC})=1^+(1^{+-})$ sector in a coupled-channels scheme, including the closest meson-meson thresholds. The meson-meson interaction is described in the framework of a constituent quark model\footnote{The interested reader is referred to Refs.~\cite{Valcarce:2005em, Segovia:2013wma} for detailed reviews about the quark model in which this work is based.} successfully employed to explain the meson and baryon phenomenology from the light to the heavy quark sector~\cite{Vijande:2004he, Valcarce:2005rr, Entem:2006dt, Valcarce:2008dr, Segovia:2008zza, Segovia:2009zz, Ortega:2011zza, Segovia:2011zza, Segovia:2015dia, Segovia:2016xqb}. Moreover, the $D^{(\ast)}\bar D^\ast$ interaction deduced from the model has been satisfactorily used to describe meson-meson~\cite{Ortega:2016mms, Ortega:2016pgg, Ortega:2017qmg} and meson-baryon~\cite{Ortega:2012cx, Ortega:2014eoa, Ortega:2014fha, Ortega:2016syt} molecular states such as the $X(3872)$ as a $D\bar D^\ast$ molecule coupled to $c\bar c(n^3P_1)$ states~\cite{Ortega:2009hj, Ortega:2012rs}.

The structure of the present manuscript is organized in the following way. In Sec.~\ref{sec:theory} the theoretical framework is briefly presented and discussed. Section~\ref{sec:results} is devoted to the analysis and discussion on the obtained results. We summarize and give some conclusions in Sec.~\ref{sec:summary}.


\section{Theoretical Formalism}
\label{sec:theory}

\subsection{Constituent quark model}

The Lagrangian of Quantum Chromodynamics (QCD) with massless light quarks is invariant under chiral rotations. This symmetry does not appear in Nature indicating that it is spontaneously broken in QCD. Among other consequences, a constituent mass which depends on the quark momentum, $M=M(q^2)$ and $M(q^2=\infty)=m_q$, is developed and Goldstone-boson exchange interactions appear between the light quarks.

Our constituent quark model (CQM) tries to mimic the previous phenomena based on the following effective Lagrangian at low-energy~\cite{Diakonov:2002fq}
\begin{equation}
{\mathcal L} = \bar{\psi}(i\, {\slash\!\!\! \partial} -M(q^{2})U^{\gamma_{5}})\,\psi  \,,
\end{equation}
being $U^{\gamma_5} = e^{i\lambda _{a}\phi ^{a}\gamma _{5}/f_{\pi}}$ the matrix of Goldstone-boson fields that can be expanded as
\begin{equation}
U^{\gamma _{5}} = 1 + \frac{i}{f_{\pi}} \gamma^{5} \lambda^{a} \pi^{a} - \frac{1}{2f_{\pi}^{2}} \pi^{a} \pi^{a} + \ldots
\end{equation}
The constituent quark mass is obtained from the first term, the second one describes the pseudoscalar meson exchange interaction among quarks and the main contribution of the third term comes from the two-pion exchange which is modeled by means of a scalar-meson exchange potential.

Another nonperturbative effect is the confining interaction which is implemented phenomenologically in order to avoid colored hadrons. In our CQM, the confinement is represented as a linear potential, due to multi-gluon exchanges between quarks, that is screened at large inter-quark distances, as a consequence of sea quarks~\cite{Bali:2005fu}:
\begin{equation}
V_{\rm CON}(\vec{r}\,)=\left[-a_{c}(1-e^{-\mu_{c}r})+\Delta \right]  (\vec{\lambda}_{q}^{c}\cdot\vec{\lambda}_{\bar{q}}^{c}) \,.
\label{eq:conf}
\end{equation}
Here, $a_{c}$ and $\mu_{c}$ are model parameters. One can see that the potential is linear at short inter-quark distances with an effective confinement strength $\sigma = -a_{c} \, \mu_{c} \, (\vec{\lambda}^{c}_{i}\cdot \vec{\lambda}^{c}_{j})$, while it becomes constant at large distances. 

Beyond the nonperturbative energy scale one expects that the dynamics of the bound-state system is governed by QCD perturbative effects. We take it into account through the one-gluon exchange potential derived from the following vertex Lagrangian
\begin{equation}
{\mathcal L}_{qqg} = i\sqrt{4\pi\alpha_{s}} \, \bar{\psi} \gamma_{\mu} 
G^{\mu}_a \lambda^a \psi,\label{Lqqg}
\end{equation}
where $\alpha_{s}$ is the strong coupling constant, $\lambda^a$ are the $SU(3)$ colour matrices and $G^{\mu}_a$ is the gluon field. To consistently treat the light- and heavy-quark sectors we employ a gluon coupling constant that scales with the reduced mass of the interacting quarks. Its explicit expression can be found in, e.g., Ref.~\cite{Vijande:2004he}.

A detailed physical background of the quark model can be found in Refs.~\cite{Vijande:2004he, Segovia:2008zz}. The model parameters and explicit expressions for the potentials can be also found therein. We want to highlight here that the interaction terms between light-light, light-heavy and heavy-heavy quarks are not the same in our formalism, i.e. while Goldstone-boson exchanges are considered when the two quarks are light, they do not appear in the other two configurations: light-heavy and heavy-heavy; however, the one-gluon exchange and confining potentials are flavor-blind. The reason is that the  Goldstone-boson exchanges originates in the spontaneous breakdown of the Chiral Symmetry.  This symmetry is explicitly broken at the level of heavy quarks and therefore no exchanges appear associated with these quarks. However, the presence of a light quark in the charm mesons allows to naturally incorporate the pion exchange interaction in the $D^{(\ast)}\bar D^\ast$ dynamics.

\subsection{Resonating Group Method and Lippmann-Schwinger equation}

The aforementioned CQM specifies the microscopic interaction among constituent quarks. In order to describe the interaction at the meson level we employ the Resonating Group Method~\cite{Wheeler:1937zza}, where mesons are considered as quark-antiquark clusters and an effective cluster-cluster interaction emerges from the underlying $q\bar q$ dynamics.

We assume that the wave function of a system composed of two mesons $A$ and $B$ with distinguishable quarks can be written as\footnote{Note that, for simplicity of the discussion presented herein, we have dropped off the spin-isospin wave function, the product of the two color singlets and the wave function that describes the center-of-mass 
motion.}
\begin{equation}
\langle \vec{p}_{A} \vec{p}_{B} \vec{P} \vec{P}_{\rm c.m.} | \psi 
\rangle = \phi_{A}(\vec{p}_{A}) \phi_{B}(\vec{p}_{B}) 
\chi_{\alpha}(\vec{P}) \,,
\label{eq:wf}
\end{equation}
where  $\phi_{C}(\vec{p}_{C})$ is the wave function of a general meson $C$ with $\vec{p}_{C}$ the relative momentum between the quark and antiquark of the meson $C$. The wave function which takes into account the relative motion of the two mesons is $\chi_\alpha(\vec{P})$, where $\alpha$ labels the set of quantum numbers needed to uniquely define a certain partial wave.

The projected Schr\"odinger equation for the relative wave function can be written as follows:
\begin{align}
&
\left(\frac{\vec{P}^{\prime 2}}{2\mu}-E \right) \chi_\alpha(\vec{P}') + \sum_{\alpha'}\int \Bigg[ {}^{\rm RGM}V_{D}^{\alpha\alpha'}(\vec{P}',\vec{P}) + \nonumber \\
&
+ {}^{\rm RGM}V_{R}^{\alpha\alpha'}(\vec{P}',\vec{P}) \Bigg] \chi_{\alpha'}(\vec{P})\, d\vec{P} = 0 \,,
\label{eq:Schrodinger}
\end{align}
where $E$ is the energy of the system. The direct potential ${}^{\rm RGM}V_{D}^{\alpha\alpha '}(\vec{P}',\vec{P})$ of a reaction $AB\to CD$ can be written as
\begin{align}
&
{}^{\rm RGM}V_{D}^{\alpha\alpha '}(\vec{P}',\vec{P}) = \sum_{i,j} \int d\vec{p}_A\, d\vec{p}_B\, d\vec{p}_C\, d\vec{p}_D\, \times \nonumber \\
&
\times \phi_C^{\ast}(\vec{p}_C) \phi_D^{\ast}(\vec{p}_D) 
V_{ij}^{\alpha\alpha '}(\vec{P}',\vec{P}) \phi_A(\vec{p}_{A}) \phi_B(\vec{p}_{B})  \,.
\end{align}
where $\{i,j\}$ runs over the constituents of the involved mesons.
The quark rearrangement potential ${}^{\rm RGM}V_{R}^{\alpha\alpha'}(\vec{P}',\vec{P})$ represents a natural way to connect meson-meson channels with different quark content, such as $\pi J/\psi$ and $D\bar D^\ast$, and it is given by
\begin{align}
&
{}^{\rm RGM}V_{R}^{\alpha\alpha'}(\vec{P}',\vec{P}) = \sum_{i,j} \int d\vec{p}_A \,
d\vec{p}_B\, d\vec{p}_C\, d\vec{p}_D\, d\vec{P^{\prime\prime}}\, \phi_{A}^{\ast}(\vec{p}_C) \times \nonumber \\
&
\times  \phi_D^{\ast}(\vec{p}_D) 
V_{ij}^{\alpha\alpha '}(\vec{P}',\vec{P}^{\prime\prime}) P_{mn} \left[\phi_A(\vec{p}_{A}) \phi_B(\vec{p}_{B}) \delta^{(3)}(\vec{P}-\vec{P}^{\prime\prime}) \right] \,,
\label{eq:Kernel}
\end{align}
where $P_{mn}$ is the operator that exchanges quarks between clusters. 

The meson eigenstates $\phi_C(\vec{p}_{C})$ are calculated by means of the two-body
Schr\"odinger equation, using the Gaussian Expansion Method~\cite{Hiyama:2003cu}. This method provides enough accuracy and simplifies the subsequent evaluation of the needed matrix elements. With the aim of optimizing the Gaussian ranges employing a reduced number of free parameters, we use Gaussian trial functions whose ranges are given by a geometrical progression~\cite{Hiyama:2003cu}. This choice produces a dense distribution at short distances enabling a better description of the dynamics mediated by short range potentials. 

The solution of the coupled-channels RGM equations is performed deriving from Eq.~\eqref{eq:Schrodinger} a set of coupled Lippmann-Schwinger equations of the form
\begin{align}
T_{\alpha}^{\alpha'}(E;p',p) &= V_{\alpha}^{\alpha'}(p',p) + \sum_{\alpha''} \int
dp''\, p^{\prime\prime2}\, V_{\alpha''}^{\alpha'}(p',p'') \nonumber \\
&
\times \frac{1}{E-{\cal E}_{\alpha''}(p^{''})}\, T_{\alpha}^{\alpha''}(E;p'',p) \,,
\end{align}
where $V_{\alpha}^{\alpha'}(p',p)$ is the projected potential that contains the direct and rearrangement potentials, and ${\cal E}_{\alpha''}(p'')$ is the energy corresponding to a momentum $p''$, written in the nonrelativistic case as:
\begin{equation}
{\cal E}_{\alpha}(p) = \frac{p^2}{2\mu_{\alpha}} + \Delta M_{\alpha} \,.
\end{equation}

Here, $\mu_{\alpha}$ is the reduced mass of the $AB$ system corresponding to the channel $\alpha$, and $\Delta M_{\alpha}$ is the difference between the threshold of the $AB$ system and the one we take as a reference.

We solve the coupled-channels Lippmann-Schwinger equation using the matrix-inversion method proposed in Ref.~\cite{Machleidt:1003bo}, generalized in order to include channels with different thresholds. Once the $T$-matrix is calculated, we determine the on-shell part which is directly related to the scattering matrix (in the case of nonrelativistic kinematics):
\begin{equation}
S_{\alpha}^{\alpha'} = 1 - 2\pi i 
\sqrt{\mu_{\alpha}\mu_{\alpha'}k_{\alpha}k_{\alpha'}} \, 
T_{\alpha}^{\alpha'}(E+i0^{+};k_{\alpha'},k_{\alpha}) \,,
\end{equation}
with $k_{\alpha}$ the on-shell momentum for channel $\alpha$.

Our aim is to explore the existence of states above and below thresholds within the same formalism. Thus, we have to analytically continue all the potentials and kernels for complex momenta in order to find the poles of the $T$-matrix in any possible Riemann sheet. 

\subsection{Production line shapes of the $Z_c$}

Most of the experimental data of the $Z_c(3900)$ and $Z_c(4020)$ were taken at $\sqrt{s}=4.26$ GeV, which corresponds to the energy of the resonance $Y(4260)$, though there are several measurements at different energies such as $\sqrt{s}=4.17$ GeV~\cite{Xiao:2013iha}, which may question the usual claim that the $Y(4260)$ is the parent of the $Z_c(3900)$. To simplify the theoretical description of the $Z_c$ generating process and to avoid possible complexities of considering the internal structure of the $Y(4260)$ resonance or other possible origin of the $Z_c(3900)$, we adopt a phenomenological vertex which creates the $\pi+A+B$ triplet, where $A+B$ is the final two-body state of the coupled-channels calculation, i.e. $A+B=\{\pi J/\psi,\, \rho\eta_c,\, D\bar D^*,\, D^\ast\bar D^\ast\}$.

The three-body decay $\pi+A+B$ of a resonance in its rest frame can be written as
\begin{equation}
 d\Gamma = \frac{(2\pi)^4}{2M}|{\cal M}|^2 d\Phi(P;p_\pi,p_A,p_B) \,,
\end{equation}
where ${\cal M}$ is the production amplitude and $d\Phi$ is the three-body phase-space given by
\begin{equation}
 d\Phi(P;p_1,p_2,p_3)=\delta^{(4)}(P-\sum_i p_i)\prod_i \frac{d^3p_i}{(2\pi)^3 2 E_i} \,.
\end{equation}
Therefore, we can express $d\Gamma$ in terms of the invariant mass spectrum of the two-meson channel, $m_{AB}$, as
\begin{equation}
d\Gamma = \frac{1}{(2\pi)^5}\frac{k_{AB} k_{\pi Z_c}}{16\,s}|{\cal M}|^2 dm_{AB} d\Omega_{AB} d\Omega_{\pi Z_c} \,,
\end{equation}
where $(k_{AB},\Omega_{AB})$ is the on-shell momentum of the $AB$ pair and $\Omega_{\pi Z_c}$ is the solid angle of the $\pi$ in the center-of-mass rest frame of $\pi+A+B$ at energy $\sqrt{s}$. The on-shell momenta are given by
\begin{eqnarray}
k_{\pi Z_c} &=& \frac{\lambda^{1/2}(\sqrt{s},m_{AB},m_\pi)}{2\sqrt{s}},\label{ec:kmomZpi}\\
k_{AB} &=& \frac{\lambda^{1/2}(m_{AB},m_A,m_B)}{2m_{AB}},\label{ec:kmomDD}
\end{eqnarray}
where $\lambda(M,m_1,m_2)=[(M^2-m_+^2)(M^2-m_-^2)]$, with $m_\pm=m_1\pm m_2$. We will assume that, in our phenomenological vertex, the production is mainly produced through the S-wave, which is expected to dominate near threshold as the D-wave is suppressed due to its momentum dependence. Integrating the angles, we have
\begin{equation}
d\Gamma = \frac{1}{(2\pi)^3}\frac{k_{AB} k_{\pi Z_c}}{4\,s}|{\cal M}^\beta(m_{AB})|^2 dm_{AB} \,,
\end{equation}
with $\beta$ the quantum numbers of the channel $AB$.

\begin{figure}[t]
\centering \includegraphics[width=1.1\columnwidth]{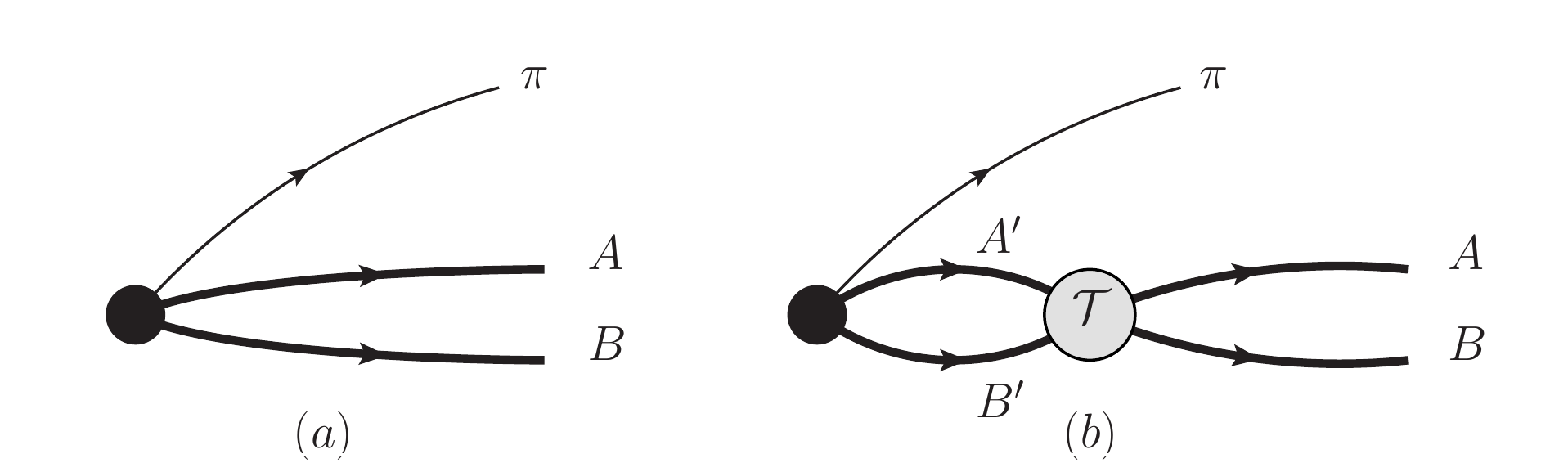}
\caption{\label{fig:Ydecay}  Background (a) and rescattering (b) contributions for the production of the two-meson $AB$ channel.}
\end{figure}

The Lorentz-invariant production amplitude, ${\cal M}$, describes the $Z_c\to AB$ production. It is diagramatically shown in Fig.~\ref{fig:Ydecay} and can be written as
\begin{equation}\label{eq:amplitude}
\mathcal{M^\beta}(m_{AB})=\left({\cal A}^\beta-\sum_{\beta'}{\cal A}^{\beta'} \int d^3p\frac{T^{\beta'\beta}(E+i0^+;p,k^\beta)}{p^2/2\mu-E-i0^+}\right).
\end{equation}
where $\beta^{(\prime)}$ denotes a given $AB$ channel in the coupled-channels calculation. We only consider production through the $AB$ in S-wave (as the vertex does not have momentum dependence) 
and we add the amplitudes ${\cal A}^\beta$ to take into account different production probabilities for each
channel $AB$.

Experimentalists measure events in the $\sigma(e^+e^-\to \pi^\pm Z_c^\mp)\times \mathcal{B}(Z_c^\mp \to AB)$ process and thus, to describe the data, we need to add a normalization factor to translate the decay rate into events:
\begin{equation}
N(m_{AB}) = \mathcal{N}_{AB}\times \frac{d\Gamma_{Z_c\to AB}}{dm_{AB}} \,.
\end{equation}

In order to fit $\{{\cal A}_{AB},\mathcal{N}_{AB}\}$ for each channel, we minimize a global $\chi^2$ function using all the available experimental data for channels $\pi J/\psi$, $D\bar D^\ast$ and $D^\ast\bar D^\ast$:
\begin{equation}
\label{eq:chisquare}
\chi^2(\{{\cal A,N}\}) = \sum_i \bigg(\frac{N^{\rm the}(x_i)-N^{\rm exp}(x_i)}{\sigma_i^{\rm exp}}\bigg)^2 \,,
\end{equation}
and the uncertainty of the parameters $\{{\cal A,N}\}$ are estimated from the second derivative of $\chi^2$ at its minimum. That is, for example, around the best-fit value ${\cal N}^{\rm BF}$, the $\chi^2$ function can be approximated by
\begin{equation}
\chi^2({\cal N})\approx \chi^2({\cal N}^{\rm BF}) + \left(\frac{{\cal N}-{\cal N}^{\rm BF}}{\sigma_{\cal N}}\right)^2 \,,
\end{equation}
with $\sigma_{\cal N}$ the error of the best-fit value ${\cal N}^{\rm BF}$.


\section{Results}
\label{sec:results}

As already mentioned above, we perform a coupled-channels calculation within the framework of the constituent quark model of Ref.~\cite{Vijande:2004he} for the $I^G(J^{PC})=1^+(1^{+-})$ sector, including the closest thresholds to the experimental masses of the $Z_c(3900)$ and $Z_c(4020)$ hadrons, that is: $\pi J/\psi$ (3234.19 MeV/$c^2$), $\rho\eta_c$ (3755.79 MeV/$c^2$), $D\bar D^\ast$ (3875.85 MeV/$c^2$), $D^\ast \bar D^\ast$ (4017.24 MeV/$c^2$), where the threshold masses are shown in parenthesis.~The $h_c\pi$ channel is not taken into account because it couples weakly to the $D^{(\ast)}\bar D^{\ast}$ channels. In fact, the only contribution of this channel would come from the $^3D_1$ component of the internal wave function of the $D^\ast$ meson, which in our model is $\sim 0.03\%$ and it is neglected in the present calculation. For the $^3S_1$ component of the $D^*$, the coupling to $h_c\pi$ is exactly zero. Similarly, we do not include other nearby channels such as the $\chi_{cJ}\rho$ ones, whose couplings are found to be almost three orders of magnitude smaller than those of the $J/\psi\pi$ and $\eta_c\rho$ channels.

\begin{figure}[!t]
\includegraphics[width=.5\textwidth]{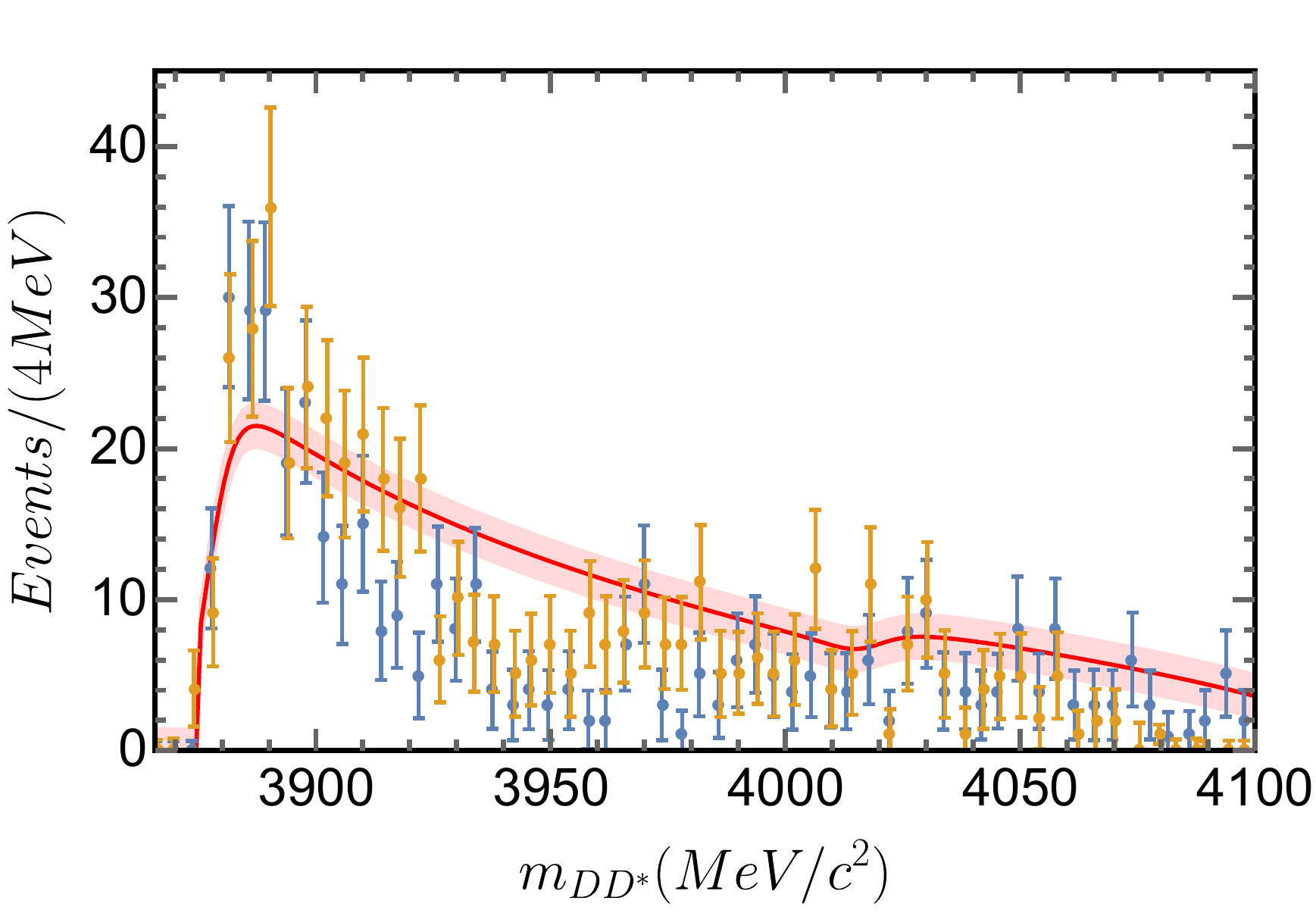}
\includegraphics[width=.5\textwidth]{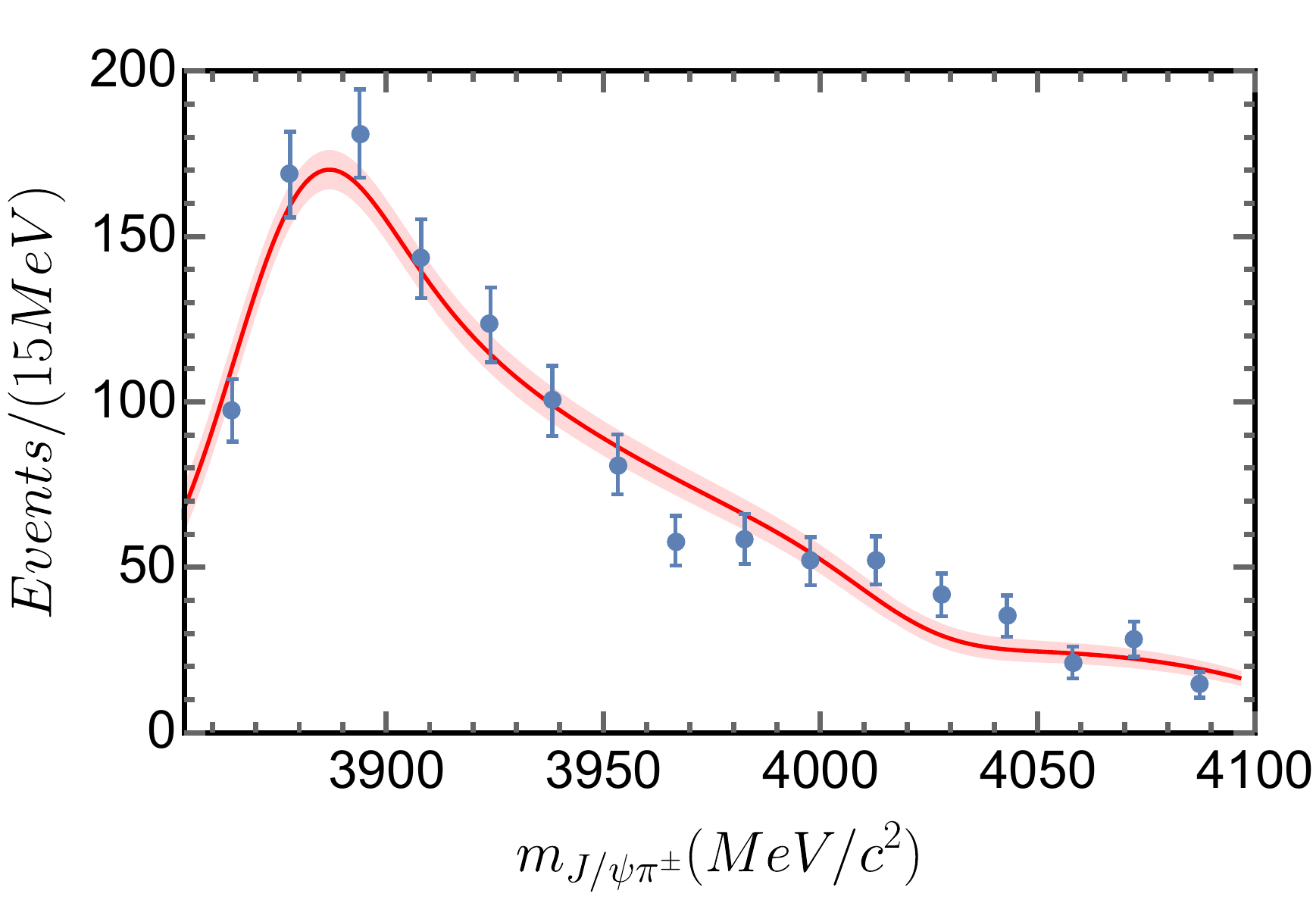}%
\caption{\label{fig:line1} Line shapes for $D\bar D^*$ (upper pannel) and $\pi J/\psi$ (lower pannel) at $\sqrt{s}=4.26$ GeV. Experimental data are from Ref.~\cite{Ablikim:2015swa,Collaboration:2017njt}, respectively. The theoretical line shapes have been convoluted with the experimental resolution. The line-shape's $68\%$ uncertainty is shown as a shadowed area.}
\end{figure}

\begin{figure}[!t]
\includegraphics[width=.5\textwidth]{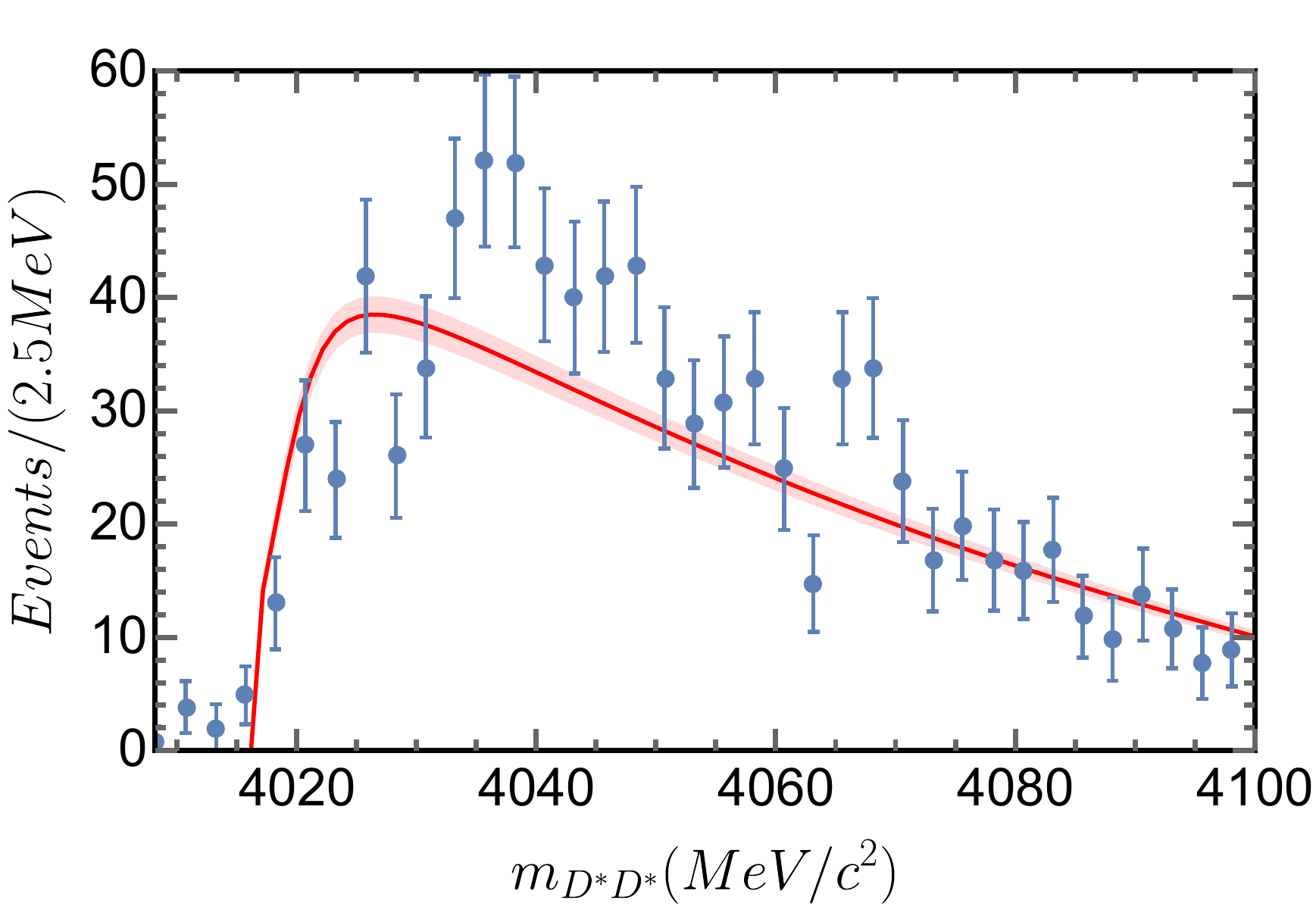}
\caption{\label{fig:line2} Line shapes for $D^*\bar D^*$ at $\sqrt{s}=4.26$ GeV. Experimental data are from Refs.~\cite{Ablikim:2013emm}. 
The theoretical line shapes have been convoluted with the experimental resolution. The line-shape's $68\%$-uncertainty is shown as a shadowed area.}
\end{figure}

Our results for the invariant mass distribution of the $D\bar D^*$, $\pi J/\psi$ and $D^*\bar D^*$ channels are shown in, respectively, the upper and lower panels of Fig.~\ref{fig:line1} and in Fig.~\ref{fig:line2}. Note that all channels mentioned in the above paragraph are included in the calculation of the theoretical line shapes. To translate the decay rates into events, we use a normalization factor ${\cal N}_{AB}$ fitted for each channel according to the expression of Eq.~\eqref{eq:chisquare}. The amplitudes ${\cal A}_{AB}$ at Eq.~\eqref{eq:amplitude} are the same for all the line shapes. The shaded area around the theoretical curve shows the statistical
68\%-confident level (CL) of the fit, obtained by propagating the errors of the fitted parameters by means of the covariance matrix. The $\chi^2$ fit is performed with the experimental results of Refs.~\cite{Ablikim:2015swa,Ablikim:2013mio,Collaboration:2017njt}, that is, experimental data for $D\bar D^*$ and $\pi J/\psi$ channels. The amplitudes ${\cal A}_{AB}$ obtained from such fit are, then, employed for the $D^*\bar D^*$ channel in order to obtain the global normalization ${\cal N}_{D^*\bar D^*}$ from experimental data of Ref~\cite{Ablikim:2013emm}. The values for the normalization factors and amplitudes are shown in Table~\ref{tab:norm}, the result on the $\chi^2/{\rm d.o.f.}$ is also collected therein. In order to describe the experimental measurement, the theoretical line shapes have been convoluted with the detector resolution.

For the fit, the $D\bar D^*$  data was taken from the $m_{D\bar D^*}$ threshold up to $3.92$ GeV. This is the region where the signal of the $Z_c(3900)$ dominates over the background~\cite{Ablikim:2013xfr}, from there on the data set is too noisy. The $\pi J/\psi$ data of Ref.~\cite{Collaboration:2017njt} was fitted from $3.85$ GeV on. In the theoretical line shape of the $D\bar D^*$ channel one can clearly see two enhancements related with the opening of the $D\bar D^*$ and $D^*\bar D^*$ thresholds, which can be associated with the $Z_c(3900)$ and the $Z_c(4020)$. In the $\pi J/\psi$ case, only one enhancement appear around $3.87$ GeV while the opening of the $D^*\bar D^*$ channel appears as a slight step down in the number of events.

\begin{table}[!t]
\begin{center}
\begin{tabular}{ccc}
\hline
\hline
Channel & $\mathcal{N}_{AB} (\times 10^{7})$ & $\mathcal{A}_{AB}$  \\
\hline
$\pi J/\psi$ & $3.76 \pm 0.09$ &  $0.34 \pm 0.01$ \\ 
$D\bar D^\ast$ & $0.80 \pm 0.04$ & $0.76  \pm 0.01$ \\
$D^\ast\bar D^\ast$ & $19.33 \pm   0.7$ & $0.66  \pm 0.01$ \\
$\rho\eta_c(1S)$ & $0.55\pm 0.10$ & $-1.00\pm0.04$ \\
\hline
$\chi_{\rm min}^2/{\rm d.o.f.}$ & \multicolumn{2}{c}{$1.89$}\\
\hline
\hline
\end{tabular}
\caption{\label{tab:norm} Production amplitudes and normalization factors for the line shapes which are fitted using Eq.~(\ref{eq:chisquare}). The minimum value of the $\chi^2/{\rm d.o.f.}$ is also given. The $68\%$-uncertainties in the amplitudes and normalization factors are obtained from the fit.}
\end{center}
\end{table}

\begin{figure}[!t]
\includegraphics[width=.5\textwidth]{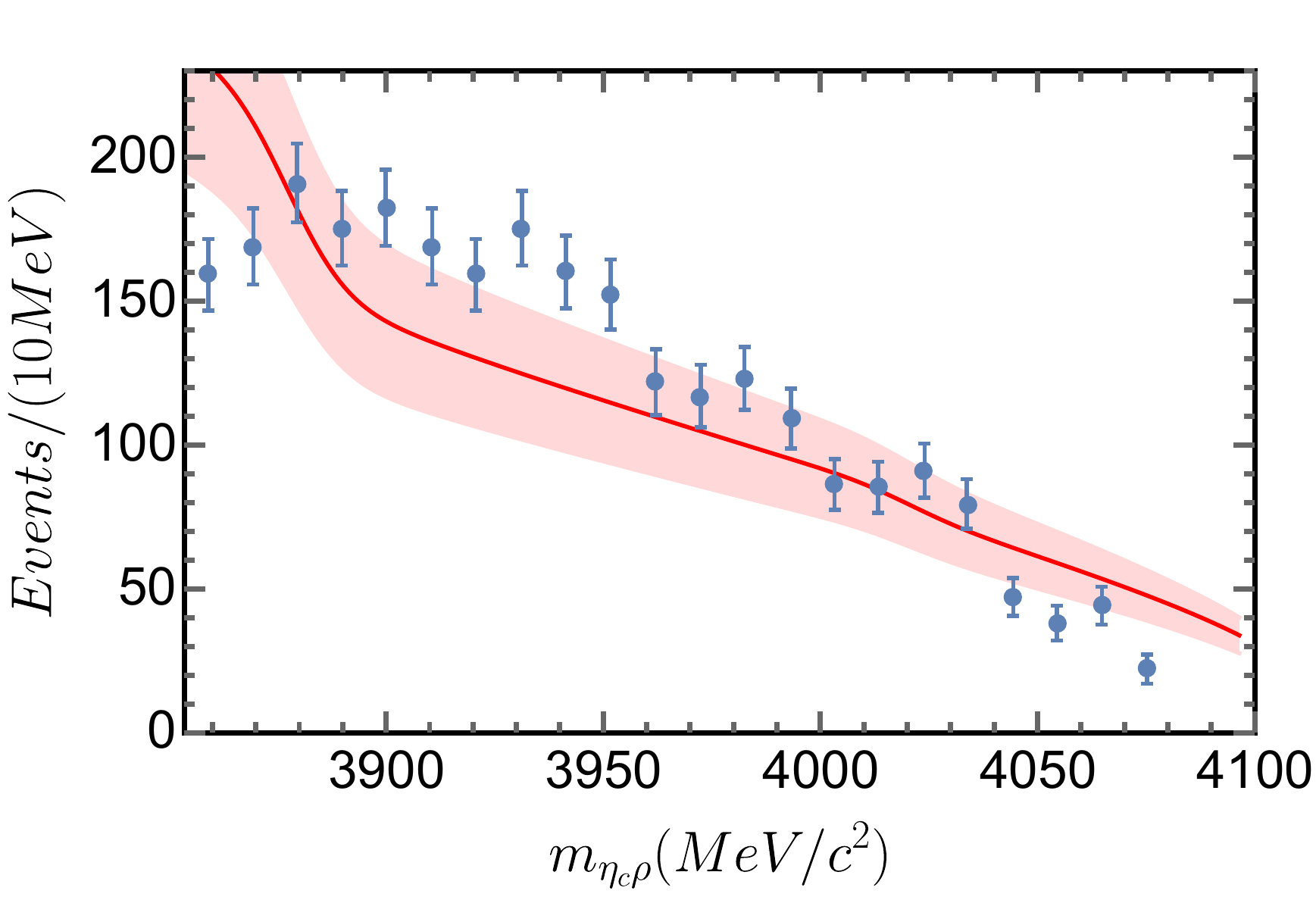}
\caption{\label{fig:line3} Line shapes for $\eta_c\rho$ channel at $\sqrt{s}=4.26$ GeV (red line). Experimental data, shown as blue dots, is taken from Ref.~\cite{Yuan:2018inv}. The theoretical line shape has been convoluted with the experimental resolution. The shadowed area shows the $68\%$ uncertainty of the line shape.}
\end{figure}

Besides the unavoidable normalization factors and amplitudes related to the phenomenological $\pi+A+B$ vertex, our formalism is able to reproduce the experimental data without fine-tuning the parameters of the constituent quark model that, in fact, are those which determine the pole position. The meson-meson interaction driven by the quark rearrangement process, which connect the $D^{(\ast)}\bar D^\ast$ channels with the $\pi J/\psi$ and $\rho\eta_c$ ones, is of the same order of magnitude than the direct interaction, $D^{(\ast)}\bar D^\ast - D^{(\ast)}\bar D^\ast$. This points to an important role of the $\pi J/\psi (\rho \eta_c)-D^{\ast}\bar D^{(\ast)}$ mixings in this sector, in agreement with the recent claims of some lattice studies~\cite{Ikeda:2016zwx}, which concluded that the $Z_c$ cannot be described as a simple $D\bar D^*$ molecule.
Recently, data on the $\rho\eta_c$ channel has been published on Ref.~\cite{Yuan:2018inv}. Although this dataset has not been included in the global fit of ${\cal A}_{AB}$ and ${\cal N}_{AB}$ parameters of Table~\ref{tab:norm}, done using solely the experimental data of $\pi J/\psi$ and $D\bar D^\ast$, we present in Fig.~\ref{fig:line3} the invariant mass distribution of the $\rho\eta_c$ channel. The ${\cal N}_{\rho\eta_c}$ global normalization is, however, adjusted from the available experimental data employing those same ${\cal A}_{AB}$ amplitudes used for all the line shapes. 

\begin{figure}[!t]
\includegraphics[width=0.5\textwidth]{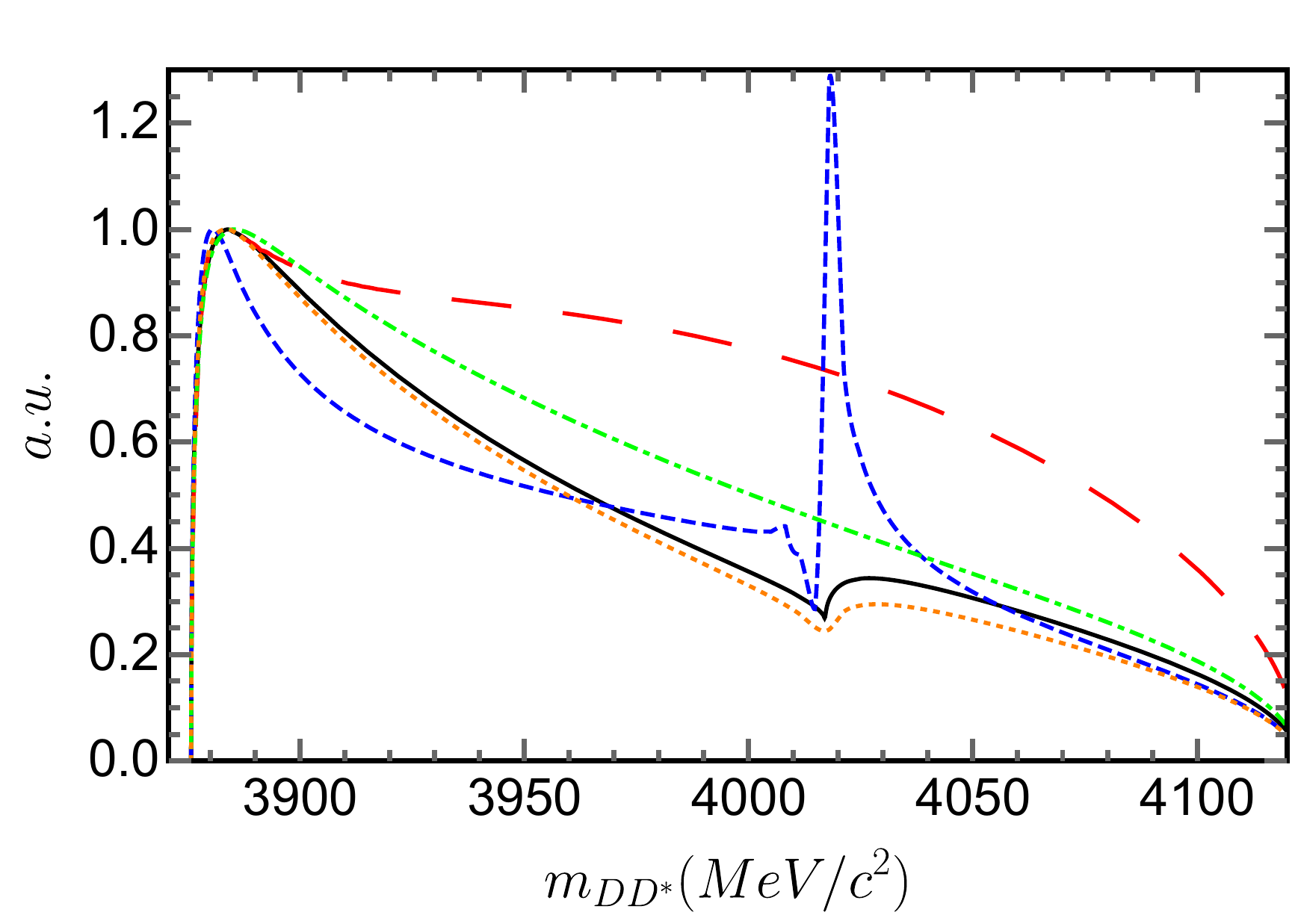}
\caption{\label{fig:test} Line shapes for different coupled-channels calculations: Only $D\bar D^*$ channel (red, long-dashed line), $D\bar D^*+D^*\bar D^*$ (blue, dashed line), $\rho\eta_c+D\bar D^\ast$ (green, dot-dashed line), $\rho\eta_c+D\bar D^\ast+D^*\bar D^*$ (orange, dotted line) and $\pi J/\psi+\rho\eta_c+D\bar D^\ast+D^*\bar D^*$ (black, solid line). All the line shapes are normalized to coincide at the $Z_c(3900)$'s peak position, according to the normalization of Table~\ref{tab:norm}. See text for discussion.}
\end{figure}

To disentangle the contribution of the different channels to the line shape, we compare in Fig.~\ref{fig:test} the line shape for the full calculation with those taking into account partial combinations of the considered channels. When the $D\bar D^\ast$ scattering channel is considered alone, a small enhancement basically due to the pion tensor interaction between the $S$ and $D$ waves appears but the generated peak is too wide. The inclusion of the $\rho\eta_c$ channel narrows that peak, making it more compatible with the experimental situation. On the other hand, adding to the $D\bar D^\ast$ channel the $D^\ast\bar D^\ast$ one generates a second structure at its threshold opening associated to the $Z_c(4020)$ peak. One can see that the second enhancement is much higher than the experimental data if we only include $D^{(\ast)}\bar D^\ast$ channels. The line shape moves closer to the experimental situation when the $\rho\eta_c$ and $\pi J/\psi$ channels are considered in the calculation, showing that they play an important role in building the observed enhancements.

To deepen into the nature of the $Z_c(3900)$ and the $Z_c(4020)$, we have examined the analytic structure on the complex energy plane of the $S$-matrix for the different coupled-channels calculations. Our results are shown in Table~\ref{tab:poles}. For the $Z_c(3900)$, one can see that even for a one-channel $D\bar D^*$ calculation the $S$-matrix shows a pole below threshold in the second Riemann sheet. When the $\rho\eta_c$ is included a pole corresponding with a virtual state appears. The inclusion of the rest of the channels does not change drastically the pole position. However, it is necessary to reproduce the experimental line shape. In the complete calculation, the $Z_c(3900)$ is associated with a pole located in the imaginary axis and the second Riemann sheet below the $D\bar D^\ast$ threshold and, thus, it is a virtual state. The situation is similar in the case of the $Z_c(4020)$, which is interpreted as a virtual state located below the $D^\ast\bar D^\ast$ threshold.

\begin{table*}[!t]
\begin{tabular}{ccccc}
\hline
\hline
Calculation & $Z_c(3900)$ pole & RS & $Z_c(4020)$ pole & RS\\
\hline
$D\bar D^*$          & $3871.37-2.17\,i$ & (S) & - & - \\ 
$D\bar D^*+D^*\bar D^*$  & $3872.27-1.85\,i$ & (S,F) & $4014.16-0.10\,i$ & (S,S) \\ 
$\rho\eta_c+D\bar D^*$  & $3871.32-0.00\,i$ & (S,S) & - & - \\  
$\rho\eta_c+D\bar D^*+D^*\bar D^*$  & $3872.07-0.00\,i$ & (S,S,F) & $4013.10-0.00\,i$ & (S,S,S) \\  
$\pi J/\psi+\rho\eta_c+D\bar D^*+D^*\bar D^*$  & $3871.74-0.00\,i$ & (S,S,S,F) & $4013.21-0.00\,i$   & (S,S,S,S) \\
\hline
\hline
\end{tabular}
\caption{\label{tab:poles} The $S$-matrix pole positions, in $\text{MeV/c}^2$, for different coupled-channels calculations. The included channels for each case are shown in the first column. Poles are given in the second and fourth columns by the value of the complex energy in a specific Riemann sheet (RS). The RS columns indicate whether the pole has been found in the first (F) or second (S) Riemann sheet of a given channel. Each channel in the coupled-channels calculation is represented as an array's element, ordered with increasing energy.}
\end{table*}

\begin{table}[!t]
\begin{tabular}{ccc}
\hline
\hline
Calculation & $Z_c(3900)$ & type \\
\hline
This work  & $3871.74$ & virtual   \\
Refs.~\cite{Aceti:2014uea} & $3878-23\,i$ & resonance  \\
Ref.~\cite{Albaladejo:2015lob} & $3894\pm6\pm1- 30\pm12\pm6\,i$ & resonance  \\
                                & $3886\pm4\pm1- 22\pm6\pm4\,i$ & resonance  \\
                                & $3831\pm 26^{+7}_{-28} $ & virtual  \\
                                & $3844\pm19^{+12}_{-21}  $ & virtual \\
Ref.~\cite{Ikeda:2016zwx} & $3709\pm 94 - 183(46)\,i$ & virtual \\
                          & $3748\pm 76 - 157(32)\,i$ & virtual \\
                          & $3686\pm 56 - 44(27)\,i$ & virtual \\
Ref.~\cite{He:2015mja}    & $3876 - 5\,i$ & resonance \\                                
\hline
Calculation & $Z_c(4020)$ & type \\
\hline
This work & $4013.21$   & virtual \\
Ref.~\cite{Aceti:2014kja} & $(3990-4000)-50\,i$ & bound/virtual \\
\hline
\hline
\end{tabular}
\caption{\label{tab:comparison} Comparison of our $S$-matrix pole position, in $\text{MeV/c}^2$, with several other calculations from literature.}
\end{table}

It is worth noticing that the real part of the $Z_c(3900)$ mass pole, shown in Table~\ref{tab:poles}, is always below the $D\bar D^*$ threshold and, hence, it does not seem to be compatible with the mass and width estimations from experimental measurements~\cite{Ablikim:2015swa, Ablikim:2013mio,Collaboration:2017njt}. Obviously, in our calculation, the $Z_c(3900)$ and $Z_c(4020)$ are not resonances but virtual state and, therefore, a directly comparison of the complex energy of the pole with the Breit-Wigner parametrization of a resonance would be misleading. The only way to connect our results with the experimental data is through the description of the line shapes.
In Tab.~\ref{tab:comparison} we show a comparison of the pole positions found in this work with other estimations from studies of the $Z_c(3900)/Z_c(4020)$ resonances.  The pole position described by other studies spans over a wide range of values in the complex plane, which could be caused by respective model details. However, as we can appreciate, most studies favour the virtual assignation, as in this work.


\section{Summary}
\label{sec:summary}
The $Z_c(3900)$ and $Z_c(4020)$ are very peculiar structures different from other molecular resonances of the charmonium spectrum. As they have $I=1$, the diagonal interaction between the $D^{(*)}\bar D^{(*)}$ channels is too suppressed to develop resonances, being the non-diagonal rearrangement interaction due to the coupling with other channels responsible for the structures appeared in the line shapes. Therefore, they should only appear in calculations involving coupled channels.

To show up this mechanism, we have performed, within the framework of a constituent quark model, a coupled-channels calculation of the $I^G(J^{PC})=1^+(1^{+-})$ sector around the energies of the $Z_c(3900)$ and $Z_c(4020)$, including the most relevant thresholds. The line shapes of the $D\bar D^*$, $\pi J/\psi$ and $D^*\bar D^*$ invariant mass distributions are well reproduced without any fine-tuning of the model parameters. The analysis of the $S$-matrix poles allows us to conclude that the point-wise behavior of the line shapes is due to the presence of two virtual states that can be seen as $D^{(*)}\bar D^*$ threshold cusps, and have an overall good agreement with the $Z_c(3900)$ and $Z_c(4020)$ signals. These results confirm the conclusion of the lattice QCD calculation of Ref.~\cite{Ikeda:2016zwx}.

\begin{acknowledgments}
This work has been partially funded by Ministerio de Econom\'ia, Industria y 
Competitividad under Contracts No. FPA2016-77177-C2-2-P, FPA2014-55613-P, FPA2017-86989-P and SEV-2016-0588 and by Junta de Castilla y Le\'on and ERDF under Contract No. SA041U16.
P.G.O. acknowledges the financial support from Spanish MINECO's Juan de la Cierva-Incorporaci\'on programme, Grant Agreement No. IJCI-2016-28525. J.S. acknowledges the financial support from the European Union's Horizon 2020 research and innovation programme under the Marie Sk\l{}odowska-Curie Grant Agreement No. 665919, and from Spanish MINECO's Juan de la Cierva-Incorporaci\'on programme, Grant Agreement No. IJCI-2016-30028.
\end{acknowledgments}


\bibliography{paperZc}

\begin{thebibliography}{72}%
\makeatletter
\providecommand \@ifxundefined [1]{%
 \@ifx{#1\undefined}
}%
\providecommand \@ifnum [1]{%
 \ifnum #1\expandafter \@firstoftwo
 \else \expandafter \@secondoftwo
 \fi
}%
\providecommand \@ifx [1]{%
 \ifx #1\expandafter \@firstoftwo
 \else \expandafter \@secondoftwo
 \fi
}%
\providecommand \natexlab [1]{#1}%
\providecommand \enquote  [1]{``#1''}%
\providecommand \bibnamefont  [1]{#1}%
\providecommand \bibfnamefont [1]{#1}%
\providecommand \citenamefont [1]{#1}%
\providecommand \href@noop [0]{\@secondoftwo}%
\providecommand \href [0]{\begingroup \@sanitize@url \@href}%
\providecommand \@href[1]{\@@startlink{#1}\@@href}%
\providecommand \@@href[1]{\endgroup#1\@@endlink}%
\providecommand \@sanitize@url [0]{\catcode `\\12\catcode `\$12\catcode
  `\&12\catcode `\#12\catcode `\^12\catcode `\_12\catcode `\%12\relax}%
\providecommand \@@startlink[1]{}%
\providecommand \@@endlink[0]{}%
\providecommand \url  [0]{\begingroup\@sanitize@url \@url }%
\providecommand \@url [1]{\endgroup\@href {#1}{\urlprefix }}%
\providecommand \urlprefix  [0]{URL }%
\providecommand \Eprint [0]{\href }%
\providecommand \doibase [0]{http://dx.doi.org/}%
\providecommand \selectlanguage [0]{\@gobble}%
\providecommand \bibinfo  [0]{\@secondoftwo}%
\providecommand \bibfield  [0]{\@secondoftwo}%
\providecommand \translation [1]{[#1]}%
\providecommand \BibitemOpen [0]{}%
\providecommand \bibitemStop [0]{}%
\providecommand \bibitemNoStop [0]{.\EOS\space}%
\providecommand \EOS [0]{\spacefactor3000\relax}%
\providecommand \BibitemShut  [1]{\csname bibitem#1\endcsname}%
\let\auto@bib@innerbib\@empty
\bibitem [{\citenamefont {Choi}\ \emph {et~al.}(2003)\citenamefont {Choi} \emph
  {et~al.}}]{Choi:2003ue}%
  \BibitemOpen
  \bibfield  {author} {\bibinfo {author} {\bibfnamefont {S.~K.}\ \bibnamefont
  {Choi}} \emph {et~al.} (\bibinfo {collaboration} {Belle}),\ }\href {\doibase
  10.1103/PhysRevLett.91.262001} {\bibfield  {journal} {\bibinfo  {journal}
  {Phys. Rev. Lett.}\ }\textbf {\bibinfo {volume} {91}},\ \bibinfo {pages}
  {262001} (\bibinfo {year} {2003})},\ \Eprint
  {http://arxiv.org/abs/hep-ex/0309032} {arXiv:hep-ex/0309032 [hep-ex]}
  \BibitemShut {NoStop}%
\bibitem [{\citenamefont {Aubert}\ \emph {et~al.}(2005)\citenamefont {Aubert}
  \emph {et~al.}}]{Aubert:2004ns}%
  \BibitemOpen
  \bibfield  {author} {\bibinfo {author} {\bibfnamefont {B.}~\bibnamefont
  {Aubert}} \emph {et~al.} (\bibinfo {collaboration} {BaBar}),\ }\href
  {\doibase 10.1103/PhysRevD.71.071103} {\bibfield  {journal} {\bibinfo
  {journal} {Phys. Rev.}\ }\textbf {\bibinfo {volume} {D71}},\ \bibinfo {pages}
  {071103} (\bibinfo {year} {2005})},\ \Eprint
  {http://arxiv.org/abs/hep-ex/0406022} {arXiv:hep-ex/0406022 [hep-ex]}
  \BibitemShut {NoStop}%
\bibitem [{\citenamefont {Acosta}\ \emph {et~al.}(2004)\citenamefont {Acosta}
  \emph {et~al.}}]{Acosta:2003zx}%
  \BibitemOpen
  \bibfield  {author} {\bibinfo {author} {\bibfnamefont {D.}~\bibnamefont
  {Acosta}} \emph {et~al.} (\bibinfo {collaboration} {CDF}),\ }\href {\doibase
  10.1103/PhysRevLett.93.072001} {\bibfield  {journal} {\bibinfo  {journal}
  {Phys. Rev. Lett.}\ }\textbf {\bibinfo {volume} {93}},\ \bibinfo {pages}
  {072001} (\bibinfo {year} {2004})},\ \Eprint
  {http://arxiv.org/abs/hep-ex/0312021} {arXiv:hep-ex/0312021 [hep-ex]}
  \BibitemShut {NoStop}%
\bibitem [{\citenamefont {Abazov}\ \emph {et~al.}(2004)\citenamefont {Abazov}
  \emph {et~al.}}]{Abazov:2004kp}%
  \BibitemOpen
  \bibfield  {author} {\bibinfo {author} {\bibfnamefont {V.~M.}\ \bibnamefont
  {Abazov}} \emph {et~al.} (\bibinfo {collaboration} {D0}),\ }\href {\doibase
  10.1103/PhysRevLett.93.162002} {\bibfield  {journal} {\bibinfo  {journal}
  {Phys. Rev. Lett.}\ }\textbf {\bibinfo {volume} {93}},\ \bibinfo {pages}
  {162002} (\bibinfo {year} {2004})},\ \Eprint
  {http://arxiv.org/abs/hep-ex/0405004} {arXiv:hep-ex/0405004 [hep-ex]}
  \BibitemShut {NoStop}%
\bibitem [{\citenamefont {Aubert}\ \emph {et~al.}(2003)\citenamefont {Aubert}
  \emph {et~al.}}]{Aubert:2003fg}%
  \BibitemOpen
  \bibfield  {author} {\bibinfo {author} {\bibfnamefont {B.}~\bibnamefont
  {Aubert}} \emph {et~al.} (\bibinfo {collaboration} {BaBar}),\ }\href
  {\doibase 10.1103/PhysRevLett.90.242001} {\bibfield  {journal} {\bibinfo
  {journal} {Phys. Rev. Lett.}\ }\textbf {\bibinfo {volume} {90}},\ \bibinfo
  {pages} {242001} (\bibinfo {year} {2003})},\ \Eprint
  {http://arxiv.org/abs/hep-ex/0304021} {arXiv:hep-ex/0304021 [hep-ex]}
  \BibitemShut {NoStop}%
\bibitem [{\citenamefont {Besson}\ \emph {et~al.}(2003)\citenamefont {Besson}
  \emph {et~al.}}]{Besson:2003cp}%
  \BibitemOpen
  \bibfield  {author} {\bibinfo {author} {\bibfnamefont {D.}~\bibnamefont
  {Besson}} \emph {et~al.} (\bibinfo {collaboration} {CLEO}),\ }\href {\doibase
  10.1103/PhysRevD.68.032002, 10.1103/PhysRevD.75.119908} {\bibfield  {journal}
  {\bibinfo  {journal} {Phys. Rev.}\ }\textbf {\bibinfo {volume} {D68}},\
  \bibinfo {pages} {032002} (\bibinfo {year} {2003})},\ \bibinfo {note}
  {[Erratum: Phys. Rev.D75,119908(2007)]},\ \Eprint
  {http://arxiv.org/abs/hep-ex/0305100} {arXiv:hep-ex/0305100 [hep-ex]}
  \BibitemShut {NoStop}%
\bibitem [{\citenamefont {Bondar}\ \emph {et~al.}(2012)\citenamefont {Bondar}
  \emph {et~al.}}]{Belle:2011aa}%
  \BibitemOpen
  \bibfield  {author} {\bibinfo {author} {\bibfnamefont {A.}~\bibnamefont
  {Bondar}} \emph {et~al.} (\bibinfo {collaboration} {Belle}),\ }\href
  {\doibase 10.1103/PhysRevLett.108.122001} {\bibfield  {journal} {\bibinfo
  {journal} {Phys. Rev. Lett.}\ }\textbf {\bibinfo {volume} {108}},\ \bibinfo
  {pages} {122001} (\bibinfo {year} {2012})},\ \Eprint
  {http://arxiv.org/abs/1110.2251} {arXiv:1110.2251 [hep-ex]} \BibitemShut
  {NoStop}%
\bibitem [{\citenamefont {Adachi}\ \emph {et~al.}(2012)\citenamefont {Adachi}
  \emph {et~al.}}]{Adachi:2012cx}%
  \BibitemOpen
  \bibfield  {author} {\bibinfo {author} {\bibfnamefont {I.}~\bibnamefont
  {Adachi}} \emph {et~al.} (\bibinfo {collaboration} {Belle})\ }(\bibinfo
  {year} {2012})\ \Eprint {http://arxiv.org/abs/1209.6450} {arXiv:1209.6450
  [hep-ex]} \BibitemShut {NoStop}%
\bibitem [{\citenamefont {Ablikim}\ \emph
  {et~al.}(2013{\natexlab{a}})\citenamefont {Ablikim} \emph
  {et~al.}}]{Ablikim:2013mio}%
  \BibitemOpen
  \bibfield  {author} {\bibinfo {author} {\bibfnamefont {M.}~\bibnamefont
  {Ablikim}} \emph {et~al.} (\bibinfo {collaboration} {BESIII}),\ }\href
  {\doibase 10.1103/PhysRevLett.110.252001} {\bibfield  {journal} {\bibinfo
  {journal} {Phys. Rev. Lett.}\ }\textbf {\bibinfo {volume} {110}},\ \bibinfo
  {pages} {252001} (\bibinfo {year} {2013}{\natexlab{a}})},\ \Eprint
  {http://arxiv.org/abs/1303.5949} {arXiv:1303.5949 [hep-ex]} \BibitemShut
  {NoStop}%
\bibitem [{\citenamefont {Liu}\ \emph {et~al.}(2013)\citenamefont {Liu} \emph
  {et~al.}}]{Liu:2013dau}%
  \BibitemOpen
  \bibfield  {author} {\bibinfo {author} {\bibfnamefont {Z.~Q.}\ \bibnamefont
  {Liu}} \emph {et~al.} (\bibinfo {collaboration} {Belle}),\ }\href {\doibase
  10.1103/PhysRevLett.110.252002} {\bibfield  {journal} {\bibinfo  {journal}
  {Phys. Rev. Lett.}\ }\textbf {\bibinfo {volume} {110}},\ \bibinfo {pages}
  {252002} (\bibinfo {year} {2013})},\ \Eprint {http://arxiv.org/abs/1304.0121}
  {arXiv:1304.0121 [hep-ex]} \BibitemShut {NoStop}%
\bibitem [{\citenamefont {Xiao}\ \emph {et~al.}(2013)\citenamefont {Xiao},
  \citenamefont {Dobbs}, \citenamefont {Tomaradze},\ and\ \citenamefont
  {Seth}}]{Xiao:2013iha}%
  \BibitemOpen
  \bibfield  {author} {\bibinfo {author} {\bibfnamefont {T.}~\bibnamefont
  {Xiao}}, \bibinfo {author} {\bibfnamefont {S.}~\bibnamefont {Dobbs}},
  \bibinfo {author} {\bibfnamefont {A.}~\bibnamefont {Tomaradze}}, \ and\
  \bibinfo {author} {\bibfnamefont {K.~K.}\ \bibnamefont {Seth}},\ }\href
  {\doibase 10.1016/j.physletb.2013.10.041} {\bibfield  {journal} {\bibinfo
  {journal} {Phys. Lett.}\ }\textbf {\bibinfo {volume} {B727}},\ \bibinfo
  {pages} {366} (\bibinfo {year} {2013})},\ \Eprint
  {http://arxiv.org/abs/1304.3036} {arXiv:1304.3036 [hep-ex]} \BibitemShut
  {NoStop}%
\bibitem [{\citenamefont {Ablikim}\ \emph
  {et~al.}(2015{\natexlab{a}})\citenamefont {Ablikim} \emph
  {et~al.}}]{Ablikim:2015tbp}%
  \BibitemOpen
  \bibfield  {author} {\bibinfo {author} {\bibfnamefont {M.}~\bibnamefont
  {Ablikim}} \emph {et~al.} (\bibinfo {collaboration} {BESIII}),\ }\href
  {\doibase 10.1103/PhysRevLett.115.112003} {\bibfield  {journal} {\bibinfo
  {journal} {Phys. Rev. Lett.}\ }\textbf {\bibinfo {volume} {115}},\ \bibinfo
  {pages} {112003} (\bibinfo {year} {2015}{\natexlab{a}})},\ \Eprint
  {http://arxiv.org/abs/1506.06018} {arXiv:1506.06018 [hep-ex]} \BibitemShut
  {NoStop}%
\bibitem [{\citenamefont {Ablikim}\ \emph
  {et~al.}(2014{\natexlab{a}})\citenamefont {Ablikim} \emph
  {et~al.}}]{Ablikim:2013xfr}%
  \BibitemOpen
  \bibfield  {author} {\bibinfo {author} {\bibfnamefont {M.}~\bibnamefont
  {Ablikim}} \emph {et~al.} (\bibinfo {collaboration} {BESIII}),\ }\href
  {\doibase 10.1103/PhysRevLett.112.022001} {\bibfield  {journal} {\bibinfo
  {journal} {Phys. Rev. Lett.}\ }\textbf {\bibinfo {volume} {112}},\ \bibinfo
  {pages} {022001} (\bibinfo {year} {2014}{\natexlab{a}})},\ \Eprint
  {http://arxiv.org/abs/1310.1163} {arXiv:1310.1163 [hep-ex]} \BibitemShut
  {NoStop}%
\bibitem [{\citenamefont {Ablikim}\ \emph
  {et~al.}(2015{\natexlab{b}})\citenamefont {Ablikim} \emph
  {et~al.}}]{Ablikim:2015swa}%
  \BibitemOpen
  \bibfield  {author} {\bibinfo {author} {\bibfnamefont {M.}~\bibnamefont
  {Ablikim}} \emph {et~al.} (\bibinfo {collaboration} {BESIII}),\ }\href
  {\doibase 10.1103/PhysRevD.92.092006} {\bibfield  {journal} {\bibinfo
  {journal} {Phys. Rev.}\ }\textbf {\bibinfo {volume} {D92}},\ \bibinfo {pages}
  {092006} (\bibinfo {year} {2015}{\natexlab{b}})},\ \Eprint
  {http://arxiv.org/abs/1509.01398} {arXiv:1509.01398 [hep-ex]} \BibitemShut
  {NoStop}%
\bibitem [{\citenamefont {Ablikim}\ \emph {et~al.}(2017)\citenamefont {Ablikim}
  \emph {et~al.}}]{Collaboration:2017njt}%
  \BibitemOpen
  \bibfield  {author} {\bibinfo {author} {\bibfnamefont {M.}~\bibnamefont
  {Ablikim}} \emph {et~al.} (\bibinfo {collaboration} {BESIII}),\ }\href
  {\doibase 10.1103/PhysRevLett.119.072001} {\bibfield  {journal} {\bibinfo
  {journal} {Phys. Rev. Lett.}\ }\textbf {\bibinfo {volume} {119}},\ \bibinfo
  {pages} {072001} (\bibinfo {year} {2017})},\ \Eprint
  {http://arxiv.org/abs/1706.04100} {arXiv:1706.04100 [hep-ex]} \BibitemShut
  {NoStop}%
\bibitem [{\citenamefont {Ablikim}\ \emph
  {et~al.}(2013{\natexlab{b}})\citenamefont {Ablikim} \emph
  {et~al.}}]{Ablikim:2013wzq}%
  \BibitemOpen
  \bibfield  {author} {\bibinfo {author} {\bibfnamefont {M.}~\bibnamefont
  {Ablikim}} \emph {et~al.} (\bibinfo {collaboration} {BESIII}),\ }\href
  {\doibase 10.1103/PhysRevLett.111.242001} {\bibfield  {journal} {\bibinfo
  {journal} {Phys. Rev. Lett.}\ }\textbf {\bibinfo {volume} {111}},\ \bibinfo
  {pages} {242001} (\bibinfo {year} {2013}{\natexlab{b}})},\ \Eprint
  {http://arxiv.org/abs/1309.1896} {arXiv:1309.1896 [hep-ex]} \BibitemShut
  {NoStop}%
\bibitem [{\citenamefont {Ablikim}\ \emph
  {et~al.}(2014{\natexlab{b}})\citenamefont {Ablikim} \emph
  {et~al.}}]{Ablikim:2013emm}%
  \BibitemOpen
  \bibfield  {author} {\bibinfo {author} {\bibfnamefont {M.}~\bibnamefont
  {Ablikim}} \emph {et~al.} (\bibinfo {collaboration} {BESIII}),\ }\href
  {\doibase 10.1103/PhysRevLett.112.132001} {\bibfield  {journal} {\bibinfo
  {journal} {Phys. Rev. Lett.}\ }\textbf {\bibinfo {volume} {112}},\ \bibinfo
  {pages} {132001} (\bibinfo {year} {2014}{\natexlab{b}})},\ \Eprint
  {http://arxiv.org/abs/1308.2760} {arXiv:1308.2760 [hep-ex]} \BibitemShut
  {NoStop}%
\bibitem [{\citenamefont {Ablikim}\ \emph
  {et~al.}(2014{\natexlab{c}})\citenamefont {Ablikim} \emph
  {et~al.}}]{Ablikim:2014dxl}%
  \BibitemOpen
  \bibfield  {author} {\bibinfo {author} {\bibfnamefont {M.}~\bibnamefont
  {Ablikim}} \emph {et~al.} (\bibinfo {collaboration} {BESIII}),\ }\href
  {\doibase 10.1103/PhysRevLett.113.212002} {\bibfield  {journal} {\bibinfo
  {journal} {Phys. Rev. Lett.}\ }\textbf {\bibinfo {volume} {113}},\ \bibinfo
  {pages} {212002} (\bibinfo {year} {2014}{\natexlab{c}})},\ \Eprint
  {http://arxiv.org/abs/1409.6577} {arXiv:1409.6577 [hep-ex]} \BibitemShut
  {NoStop}%
\bibitem [{\citenamefont {Guo}\ \emph {et~al.}(2013)\citenamefont {Guo},
  \citenamefont {Hidalgo-Duque}, \citenamefont {Nieves},\ and\ \citenamefont
  {Valderrama}}]{Guo:2013sya}%
  \BibitemOpen
  \bibfield  {author} {\bibinfo {author} {\bibfnamefont {F.-K.}\ \bibnamefont
  {Guo}}, \bibinfo {author} {\bibfnamefont {C.}~\bibnamefont {Hidalgo-Duque}},
  \bibinfo {author} {\bibfnamefont {J.}~\bibnamefont {Nieves}}, \ and\ \bibinfo
  {author} {\bibfnamefont {M.~P.}\ \bibnamefont {Valderrama}},\ }\href
  {\doibase 10.1103/PhysRevD.88.054007} {\bibfield  {journal} {\bibinfo
  {journal} {Phys. Rev.}\ }\textbf {\bibinfo {volume} {D88}},\ \bibinfo {pages}
  {054007} (\bibinfo {year} {2013})},\ \Eprint {http://arxiv.org/abs/1303.6608}
  {arXiv:1303.6608 [hep-ph]} \BibitemShut {NoStop}%
\bibitem [{\citenamefont {Mehen}\ and\ \citenamefont
  {Powell}(2013)}]{Mehen:2013mva}%
  \BibitemOpen
  \bibfield  {author} {\bibinfo {author} {\bibfnamefont {T.}~\bibnamefont
  {Mehen}}\ and\ \bibinfo {author} {\bibfnamefont {J.}~\bibnamefont {Powell}},\
  }\href {\doibase 10.1103/PhysRevD.88.034017} {\bibfield  {journal} {\bibinfo
  {journal} {Phys. Rev.}\ }\textbf {\bibinfo {volume} {D88}},\ \bibinfo {pages}
  {034017} (\bibinfo {year} {2013})},\ \Eprint {http://arxiv.org/abs/1306.5459}
  {arXiv:1306.5459 [hep-ph]} \BibitemShut {NoStop}%
\bibitem [{\citenamefont {He}\ \emph {et~al.}(2013)\citenamefont {He},
  \citenamefont {Liu}, \citenamefont {Sun},\ and\ \citenamefont
  {Zhu}}]{He:2013nwa}%
  \BibitemOpen
  \bibfield  {author} {\bibinfo {author} {\bibfnamefont {J.}~\bibnamefont
  {He}}, \bibinfo {author} {\bibfnamefont {X.}~\bibnamefont {Liu}}, \bibinfo
  {author} {\bibfnamefont {Z.-F.}\ \bibnamefont {Sun}}, \ and\ \bibinfo
  {author} {\bibfnamefont {S.-L.}\ \bibnamefont {Zhu}},\ }\href {\doibase
  10.1140/epjc/s10052-013-2635-z} {\bibfield  {journal} {\bibinfo  {journal}
  {Eur. Phys. J.}\ }\textbf {\bibinfo {volume} {C73}},\ \bibinfo {pages} {2635}
  (\bibinfo {year} {2013})},\ \Eprint {http://arxiv.org/abs/1308.2999}
  {arXiv:1308.2999 [hep-ph]} \BibitemShut {NoStop}%
\bibitem [{\citenamefont {Liu}\ \emph {et~al.}(2014)\citenamefont {Liu},
  \citenamefont {Ma}, \citenamefont {Sun}, \citenamefont {Liu},\ and\
  \citenamefont {Zhu}}]{Liu:2014eka}%
  \BibitemOpen
  \bibfield  {author} {\bibinfo {author} {\bibfnamefont {X.-H.}\ \bibnamefont
  {Liu}}, \bibinfo {author} {\bibfnamefont {L.}~\bibnamefont {Ma}}, \bibinfo
  {author} {\bibfnamefont {L.-P.}\ \bibnamefont {Sun}}, \bibinfo {author}
  {\bibfnamefont {X.}~\bibnamefont {Liu}}, \ and\ \bibinfo {author}
  {\bibfnamefont {S.-L.}\ \bibnamefont {Zhu}},\ }\href {\doibase
  10.1103/PhysRevD.90.074020} {\bibfield  {journal} {\bibinfo  {journal} {Phys.
  Rev.}\ }\textbf {\bibinfo {volume} {D90}},\ \bibinfo {pages} {074020}
  (\bibinfo {year} {2014})},\ \Eprint {http://arxiv.org/abs/1407.3684}
  {arXiv:1407.3684 [hep-ph]} \BibitemShut {NoStop}%
\bibitem [{\citenamefont {Esposito}\ \emph {et~al.}(2015)\citenamefont
  {Esposito}, \citenamefont {Guerrieri}, \citenamefont {Piccinini},
  \citenamefont {Pilloni},\ and\ \citenamefont {Polosa}}]{Esposito:2014rxa}%
  \BibitemOpen
  \bibfield  {author} {\bibinfo {author} {\bibfnamefont {A.}~\bibnamefont
  {Esposito}}, \bibinfo {author} {\bibfnamefont {A.~L.}\ \bibnamefont
  {Guerrieri}}, \bibinfo {author} {\bibfnamefont {F.}~\bibnamefont
  {Piccinini}}, \bibinfo {author} {\bibfnamefont {A.}~\bibnamefont {Pilloni}},
  \ and\ \bibinfo {author} {\bibfnamefont {A.~D.}\ \bibnamefont {Polosa}},\
  }\href {\doibase 10.1142/S0217751X15300021} {\bibfield  {journal} {\bibinfo
  {journal} {Int. J. Mod. Phys.}\ }\textbf {\bibinfo {volume} {A30}},\ \bibinfo
  {pages} {1530002} (\bibinfo {year} {2015})},\ \Eprint
  {http://arxiv.org/abs/1411.5997} {arXiv:1411.5997 [hep-ph]} \BibitemShut
  {NoStop}%
\bibitem [{\citenamefont {Dias}\ \emph {et~al.}(2013)\citenamefont {Dias},
  \citenamefont {Navarra}, \citenamefont {Nielsen},\ and\ \citenamefont
  {Zanetti}}]{Dias:2013xfa}%
  \BibitemOpen
  \bibfield  {author} {\bibinfo {author} {\bibfnamefont {J.~M.}\ \bibnamefont
  {Dias}}, \bibinfo {author} {\bibfnamefont {F.~S.}\ \bibnamefont {Navarra}},
  \bibinfo {author} {\bibfnamefont {M.}~\bibnamefont {Nielsen}}, \ and\
  \bibinfo {author} {\bibfnamefont {C.~M.}\ \bibnamefont {Zanetti}},\ }\href
  {\doibase 10.1103/PhysRevD.88.016004} {\bibfield  {journal} {\bibinfo
  {journal} {Phys. Rev.}\ }\textbf {\bibinfo {volume} {D88}},\ \bibinfo {pages}
  {016004} (\bibinfo {year} {2013})},\ \Eprint {http://arxiv.org/abs/1304.6433}
  {arXiv:1304.6433 [hep-ph]} \BibitemShut {NoStop}%
\bibitem [{\citenamefont {Agaev}\ \emph {et~al.}(2017)\citenamefont {Agaev},
  \citenamefont {Azizi},\ and\ \citenamefont {Sundu}}]{Agaev:2017tzv}%
  \BibitemOpen
  \bibfield  {author} {\bibinfo {author} {\bibfnamefont {S.~S.}\ \bibnamefont
  {Agaev}}, \bibinfo {author} {\bibfnamefont {K.}~\bibnamefont {Azizi}}, \ and\
  \bibinfo {author} {\bibfnamefont {H.}~\bibnamefont {Sundu}},\ }\href
  {\doibase 10.1103/PhysRevD.96.034026} {\bibfield  {journal} {\bibinfo
  {journal} {Phys. Rev.}\ }\textbf {\bibinfo {volume} {D96}},\ \bibinfo {pages}
  {034026} (\bibinfo {year} {2017})},\ \Eprint
  {http://arxiv.org/abs/1706.01216} {arXiv:1706.01216 [hep-ph]} \BibitemShut
  {NoStop}%
\bibitem [{\citenamefont {Wang}\ and\ \citenamefont
  {Huang}(2014)}]{Wang:2013vex}%
  \BibitemOpen
  \bibfield  {author} {\bibinfo {author} {\bibfnamefont {Z.-G.}\ \bibnamefont
  {Wang}}\ and\ \bibinfo {author} {\bibfnamefont {T.}~\bibnamefont {Huang}},\
  }\href {\doibase 10.1103/PhysRevD.89.054019} {\bibfield  {journal} {\bibinfo
  {journal} {Phys. Rev.}\ }\textbf {\bibinfo {volume} {D89}},\ \bibinfo {pages}
  {054019} (\bibinfo {year} {2014})},\ \Eprint {http://arxiv.org/abs/1310.2422}
  {arXiv:1310.2422 [hep-ph]} \BibitemShut {NoStop}%
\bibitem [{\citenamefont {Qiao}\ and\ \citenamefont
  {Tang}(2014)}]{Qiao:2013raa}%
  \BibitemOpen
  \bibfield  {author} {\bibinfo {author} {\bibfnamefont {C.-F.}\ \bibnamefont
  {Qiao}}\ and\ \bibinfo {author} {\bibfnamefont {L.}~\bibnamefont {Tang}},\
  }\href {\doibase 10.1140/epjc/s10052-014-3122-x} {\bibfield  {journal}
  {\bibinfo  {journal} {Eur. Phys. J.}\ }\textbf {\bibinfo {volume} {C74}},\
  \bibinfo {pages} {3122} (\bibinfo {year} {2014})},\ \Eprint
  {http://arxiv.org/abs/1307.6654} {arXiv:1307.6654 [hep-ph]} \BibitemShut
  {NoStop}%
\bibitem [{\citenamefont {Deng}\ \emph {et~al.}(2014)\citenamefont {Deng},
  \citenamefont {Ping},\ and\ \citenamefont {Wang}}]{Deng:2014gqa}%
  \BibitemOpen
  \bibfield  {author} {\bibinfo {author} {\bibfnamefont {C.}~\bibnamefont
  {Deng}}, \bibinfo {author} {\bibfnamefont {J.}~\bibnamefont {Ping}}, \ and\
  \bibinfo {author} {\bibfnamefont {F.}~\bibnamefont {Wang}},\ }\href {\doibase
  10.1103/PhysRevD.90.054009} {\bibfield  {journal} {\bibinfo  {journal} {Phys.
  Rev.}\ }\textbf {\bibinfo {volume} {D90}},\ \bibinfo {pages} {054009}
  (\bibinfo {year} {2014})},\ \Eprint {http://arxiv.org/abs/1402.0777}
  {arXiv:1402.0777 [hep-ph]} \BibitemShut {NoStop}%
\bibitem [{\citenamefont {Swanson}(2015)}]{Swanson:2014tra}%
  \BibitemOpen
  \bibfield  {author} {\bibinfo {author} {\bibfnamefont {E.~S.}\ \bibnamefont
  {Swanson}},\ }\href {\doibase 10.1103/PhysRevD.91.034009} {\bibfield
  {journal} {\bibinfo  {journal} {Phys. Rev.}\ }\textbf {\bibinfo {volume}
  {D91}},\ \bibinfo {pages} {034009} (\bibinfo {year} {2015})},\ \Eprint
  {http://arxiv.org/abs/1409.3291} {arXiv:1409.3291 [hep-ph]} \BibitemShut
  {NoStop}%
\bibitem [{\citenamefont {Swanson}(2016)}]{Swanson:2015bsa}%
  \BibitemOpen
  \bibfield  {author} {\bibinfo {author} {\bibfnamefont {E.~S.}\ \bibnamefont
  {Swanson}},\ }\href {\doibase 10.1142/S0218301316420106} {\bibfield
  {journal} {\bibinfo  {journal} {Int. J. Mod. Phys.}\ }\textbf {\bibinfo
  {volume} {E25}},\ \bibinfo {pages} {1642010} (\bibinfo {year} {2016})},\
  \Eprint {http://arxiv.org/abs/1504.07952} {arXiv:1504.07952 [hep-ph]}
  \BibitemShut {NoStop}%
\bibitem [{\citenamefont {Newton}(1982)}]{Newton:1982qc}%
  \BibitemOpen
  \bibfield  {author} {\bibinfo {author} {\bibfnamefont {R.~G.}\ \bibnamefont
  {Newton}},\ }\href@noop {} {\emph {\bibinfo {title} {{SCATTERING THEORY OF
  WAVES AND PARTICLES}}}}\ (\bibinfo {year} {1982})\BibitemShut {NoStop}%
\bibitem [{\citenamefont {Guo}\ \emph {et~al.}(2015)\citenamefont {Guo},
  \citenamefont {Hanhart}, \citenamefont {Wang},\ and\ \citenamefont
  {Zhao}}]{Guo:2014iya}%
  \BibitemOpen
  \bibfield  {author} {\bibinfo {author} {\bibfnamefont {F.-K.}\ \bibnamefont
  {Guo}}, \bibinfo {author} {\bibfnamefont {C.}~\bibnamefont {Hanhart}},
  \bibinfo {author} {\bibfnamefont {Q.}~\bibnamefont {Wang}}, \ and\ \bibinfo
  {author} {\bibfnamefont {Q.}~\bibnamefont {Zhao}},\ }\href {\doibase
  10.1103/PhysRevD.91.051504} {\bibfield  {journal} {\bibinfo  {journal} {Phys.
  Rev.}\ }\textbf {\bibinfo {volume} {D91}},\ \bibinfo {pages} {051504}
  (\bibinfo {year} {2015})},\ \Eprint {http://arxiv.org/abs/1411.5584}
  {arXiv:1411.5584 [hep-ph]} \BibitemShut {NoStop}%
\bibitem [{\citenamefont {Pilloni}\ \emph {et~al.}(2017)\citenamefont
  {Pilloni}, \citenamefont {Fernandez-Ramirez}, \citenamefont {Jackura},
  \citenamefont {Mathieu}, \citenamefont {Mikhasenko}, \citenamefont {Nys},\
  and\ \citenamefont {Szczepaniak}}]{Pilloni:2016obd}%
  \BibitemOpen
  \bibfield  {author} {\bibinfo {author} {\bibfnamefont {A.}~\bibnamefont
  {Pilloni}}, \bibinfo {author} {\bibfnamefont {C.}~\bibnamefont
  {Fernandez-Ramirez}}, \bibinfo {author} {\bibfnamefont {A.}~\bibnamefont
  {Jackura}}, \bibinfo {author} {\bibfnamefont {V.}~\bibnamefont {Mathieu}},
  \bibinfo {author} {\bibfnamefont {M.}~\bibnamefont {Mikhasenko}}, \bibinfo
  {author} {\bibfnamefont {J.}~\bibnamefont {Nys}}, \ and\ \bibinfo {author}
  {\bibfnamefont {A.~P.}\ \bibnamefont {Szczepaniak}} (\bibinfo {collaboration}
  {JPAC}),\ }\href {\doibase 10.1016/j.physletb.2017.06.030} {\bibfield
  {journal} {\bibinfo  {journal} {Phys. Lett.}\ }\textbf {\bibinfo {volume}
  {B772}},\ \bibinfo {pages} {200} (\bibinfo {year} {2017})},\ \Eprint
  {http://arxiv.org/abs/1612.06490} {arXiv:1612.06490 [hep-ph]} \BibitemShut
  {NoStop}%
\bibitem [{\citenamefont {Aceti}\ \emph
  {et~al.}(2014{\natexlab{a}})\citenamefont {Aceti}, \citenamefont {Bayar},
  \citenamefont {Oset}, \citenamefont {Martinez~Torres}, \citenamefont
  {Khemchandani}, \citenamefont {Dias}, \citenamefont {Navarra},\ and\
  \citenamefont {Nielsen}}]{Aceti:2014uea}%
  \BibitemOpen
  \bibfield  {author} {\bibinfo {author} {\bibfnamefont {F.}~\bibnamefont
  {Aceti}}, \bibinfo {author} {\bibfnamefont {M.}~\bibnamefont {Bayar}},
  \bibinfo {author} {\bibfnamefont {E.}~\bibnamefont {Oset}}, \bibinfo {author}
  {\bibfnamefont {A.}~\bibnamefont {Martinez~Torres}}, \bibinfo {author}
  {\bibfnamefont {K.~P.}\ \bibnamefont {Khemchandani}}, \bibinfo {author}
  {\bibfnamefont {J.~M.}\ \bibnamefont {Dias}}, \bibinfo {author}
  {\bibfnamefont {F.~S.}\ \bibnamefont {Navarra}}, \ and\ \bibinfo {author}
  {\bibfnamefont {M.}~\bibnamefont {Nielsen}},\ }\href {\doibase
  10.1103/PhysRevD.90.016003} {\bibfield  {journal} {\bibinfo  {journal} {Phys.
  Rev.}\ }\textbf {\bibinfo {volume} {D90}},\ \bibinfo {pages} {016003}
  (\bibinfo {year} {2014}{\natexlab{a}})},\ \Eprint
  {http://arxiv.org/abs/1401.8216} {arXiv:1401.8216 [hep-ph]} \BibitemShut
  {NoStop}%
\bibitem [{\citenamefont {Albaladejo}\ \emph {et~al.}(2016)\citenamefont
  {Albaladejo}, \citenamefont {Guo}, \citenamefont {Hidalgo-Duque},\ and\
  \citenamefont {Nieves}}]{Albaladejo:2015lob}%
  \BibitemOpen
  \bibfield  {author} {\bibinfo {author} {\bibfnamefont {M.}~\bibnamefont
  {Albaladejo}}, \bibinfo {author} {\bibfnamefont {F.-K.}\ \bibnamefont {Guo}},
  \bibinfo {author} {\bibfnamefont {C.}~\bibnamefont {Hidalgo-Duque}}, \ and\
  \bibinfo {author} {\bibfnamefont {J.}~\bibnamefont {Nieves}},\ }\href
  {\doibase 10.1016/j.physletb.2016.02.025} {\bibfield  {journal} {\bibinfo
  {journal} {Phys. Lett.}\ }\textbf {\bibinfo {volume} {B755}},\ \bibinfo
  {pages} {337} (\bibinfo {year} {2016})},\ \Eprint
  {http://arxiv.org/abs/1512.03638} {arXiv:1512.03638 [hep-ph]} \BibitemShut
  {NoStop}%
\bibitem [{\citenamefont {He}\ and\ \citenamefont {Chen}(2018)}]{He:2017lhy}%
  \BibitemOpen
  \bibfield  {author} {\bibinfo {author} {\bibfnamefont {J.}~\bibnamefont
  {He}}\ and\ \bibinfo {author} {\bibfnamefont {D.-Y.}\ \bibnamefont {Chen}},\
  }\href {\doibase 10.1140/epjc/s10052-018-5580-z} {\bibfield  {journal}
  {\bibinfo  {journal} {Eur. Phys. J.}\ }\textbf {\bibinfo {volume} {C78}},\
  \bibinfo {pages} {94} (\bibinfo {year} {2018})},\ \Eprint
  {http://arxiv.org/abs/1712.05653} {arXiv:1712.05653 [hep-ph]} \BibitemShut
  {NoStop}%
\bibitem [{\citenamefont {Aceti}\ \emph
  {et~al.}(2014{\natexlab{b}})\citenamefont {Aceti}, \citenamefont {Bayar},
  \citenamefont {Dias},\ and\ \citenamefont {Oset}}]{Aceti:2014kja}%
  \BibitemOpen
  \bibfield  {author} {\bibinfo {author} {\bibfnamefont {F.}~\bibnamefont
  {Aceti}}, \bibinfo {author} {\bibfnamefont {M.}~\bibnamefont {Bayar}},
  \bibinfo {author} {\bibfnamefont {J.~M.}\ \bibnamefont {Dias}}, \ and\
  \bibinfo {author} {\bibfnamefont {E.}~\bibnamefont {Oset}},\ }\href {\doibase
  10.1140/epja/i2014-14103-1} {\bibfield  {journal} {\bibinfo  {journal} {Eur.
  Phys. J.}\ }\textbf {\bibinfo {volume} {A50}},\ \bibinfo {pages} {103}
  (\bibinfo {year} {2014}{\natexlab{b}})},\ \Eprint
  {http://arxiv.org/abs/1401.2076} {arXiv:1401.2076 [hep-ph]} \BibitemShut
  {NoStop}%
\bibitem [{\citenamefont {Prelovsek}\ \emph {et~al.}(2015)\citenamefont
  {Prelovsek}, \citenamefont {Lang}, \citenamefont {Leskovec},\ and\
  \citenamefont {Mohler}}]{Prelovsek:2014swa}%
  \BibitemOpen
  \bibfield  {author} {\bibinfo {author} {\bibfnamefont {S.}~\bibnamefont
  {Prelovsek}}, \bibinfo {author} {\bibfnamefont {C.~B.}\ \bibnamefont {Lang}},
  \bibinfo {author} {\bibfnamefont {L.}~\bibnamefont {Leskovec}}, \ and\
  \bibinfo {author} {\bibfnamefont {D.}~\bibnamefont {Mohler}},\ }\href
  {\doibase 10.1103/PhysRevD.91.014504} {\bibfield  {journal} {\bibinfo
  {journal} {Phys. Rev.}\ }\textbf {\bibinfo {volume} {D91}},\ \bibinfo {pages}
  {014504} (\bibinfo {year} {2015})},\ \Eprint {http://arxiv.org/abs/1405.7623}
  {arXiv:1405.7623 [hep-lat]} \BibitemShut {NoStop}%
\bibitem [{\citenamefont {Prelovsek}\ and\ \citenamefont
  {Leskovec}(2013)}]{Prelovsek:2013xba}%
  \BibitemOpen
  \bibfield  {author} {\bibinfo {author} {\bibfnamefont {S.}~\bibnamefont
  {Prelovsek}}\ and\ \bibinfo {author} {\bibfnamefont {L.}~\bibnamefont
  {Leskovec}},\ }\href {\doibase 10.1016/j.physletb.2013.10.009} {\bibfield
  {journal} {\bibinfo  {journal} {Phys. Lett.}\ }\textbf {\bibinfo {volume}
  {B727}},\ \bibinfo {pages} {172} (\bibinfo {year} {2013})},\ \Eprint
  {http://arxiv.org/abs/1308.2097} {arXiv:1308.2097 [hep-lat]} \BibitemShut
  {NoStop}%
\bibitem [{\citenamefont {Chen}\ \emph {et~al.}(2014)\citenamefont {Chen} \emph
  {et~al.}}]{Chen:2014afa}%
  \BibitemOpen
  \bibfield  {author} {\bibinfo {author} {\bibfnamefont {Y.}~\bibnamefont
  {Chen}} \emph {et~al.},\ }\href {\doibase 10.1103/PhysRevD.89.094506}
  {\bibfield  {journal} {\bibinfo  {journal} {Phys. Rev.}\ }\textbf {\bibinfo
  {volume} {D89}},\ \bibinfo {pages} {094506} (\bibinfo {year} {2014})},\
  \Eprint {http://arxiv.org/abs/1403.1318} {arXiv:1403.1318 [hep-lat]}
  \BibitemShut {NoStop}%
\bibitem [{\citenamefont {Chen}\ \emph {et~al.}(2015)\citenamefont {Chen} \emph
  {et~al.}}]{Chen:2015jwa}%
  \BibitemOpen
  \bibfield  {author} {\bibinfo {author} {\bibfnamefont {Y.}~\bibnamefont
  {Chen}} \emph {et~al.} (\bibinfo {collaboration} {CLQCD}),\ }\href {\doibase
  10.1103/PhysRevD.92.054507} {\bibfield  {journal} {\bibinfo  {journal} {Phys.
  Rev.}\ }\textbf {\bibinfo {volume} {D92}},\ \bibinfo {pages} {054507}
  (\bibinfo {year} {2015})},\ \Eprint {http://arxiv.org/abs/1503.02371}
  {arXiv:1503.02371 [hep-lat]} \BibitemShut {NoStop}%
\bibitem [{\citenamefont {Lee}\ \emph {et~al.}(2014)\citenamefont {Lee},
  \citenamefont {DeTar}, \citenamefont {Na},\ and\ \citenamefont
  {Mohler}}]{Lee:2014uta}%
  \BibitemOpen
  \bibfield  {author} {\bibinfo {author} {\bibfnamefont {S.-h.}\ \bibnamefont
  {Lee}}, \bibinfo {author} {\bibfnamefont {C.}~\bibnamefont {DeTar}}, \bibinfo
  {author} {\bibfnamefont {H.}~\bibnamefont {Na}}, \ and\ \bibinfo {author}
  {\bibfnamefont {D.}~\bibnamefont {Mohler}} (\bibinfo {collaboration}
  {Fermilab Lattice, MILC}),\ }\href@noop {} {\  (\bibinfo {year} {2014})},\
  \Eprint {http://arxiv.org/abs/1411.1389} {arXiv:1411.1389 [hep-lat]}
  \BibitemShut {NoStop}%
\bibitem [{\citenamefont {Ikeda}\ \emph {et~al.}(2016)\citenamefont {Ikeda},
  \citenamefont {Aoki}, \citenamefont {Doi}, \citenamefont {Gongyo},
  \citenamefont {Hatsuda}, \citenamefont {Inoue}, \citenamefont {Iritani},
  \citenamefont {Ishii}, \citenamefont {Murano},\ and\ \citenamefont
  {Sasaki}}]{Ikeda:2016zwx}%
  \BibitemOpen
  \bibfield  {author} {\bibinfo {author} {\bibfnamefont {Y.}~\bibnamefont
  {Ikeda}}, \bibinfo {author} {\bibfnamefont {S.}~\bibnamefont {Aoki}},
  \bibinfo {author} {\bibfnamefont {T.}~\bibnamefont {Doi}}, \bibinfo {author}
  {\bibfnamefont {S.}~\bibnamefont {Gongyo}}, \bibinfo {author} {\bibfnamefont
  {T.}~\bibnamefont {Hatsuda}}, \bibinfo {author} {\bibfnamefont
  {T.}~\bibnamefont {Inoue}}, \bibinfo {author} {\bibfnamefont
  {T.}~\bibnamefont {Iritani}}, \bibinfo {author} {\bibfnamefont
  {N.}~\bibnamefont {Ishii}}, \bibinfo {author} {\bibfnamefont
  {K.}~\bibnamefont {Murano}}, \ and\ \bibinfo {author} {\bibfnamefont
  {K.}~\bibnamefont {Sasaki}} (\bibinfo {collaboration} {HAL QCD}),\ }\href
  {\doibase 10.1103/PhysRevLett.117.242001} {\bibfield  {journal} {\bibinfo
  {journal} {Phys. Rev. Lett.}\ }\textbf {\bibinfo {volume} {117}},\ \bibinfo
  {pages} {242001} (\bibinfo {year} {2016})},\ \Eprint
  {http://arxiv.org/abs/1602.03465} {arXiv:1602.03465 [hep-lat]} \BibitemShut
  {NoStop}%
\bibitem [{\citenamefont {Valcarce}\ \emph
  {et~al.}(2005{\natexlab{a}})\citenamefont {Valcarce}, \citenamefont
  {Garcilazo}, \citenamefont {Fernandez},\ and\ \citenamefont
  {Gonzalez}}]{Valcarce:2005em}%
  \BibitemOpen
  \bibfield  {author} {\bibinfo {author} {\bibfnamefont {A.}~\bibnamefont
  {Valcarce}}, \bibinfo {author} {\bibfnamefont {H.}~\bibnamefont {Garcilazo}},
  \bibinfo {author} {\bibfnamefont {F.}~\bibnamefont {Fernandez}}, \ and\
  \bibinfo {author} {\bibfnamefont {P.}~\bibnamefont {Gonzalez}},\ }\href
  {\doibase 10.1088/0034-4885/68/5/R01} {\bibfield  {journal} {\bibinfo
  {journal} {Rept. Prog. Phys.}\ }\textbf {\bibinfo {volume} {68}},\ \bibinfo
  {pages} {965} (\bibinfo {year} {2005}{\natexlab{a}})},\ \Eprint
  {http://arxiv.org/abs/hep-ph/0502173} {arXiv:hep-ph/0502173 [hep-ph]}
  \BibitemShut {NoStop}%
\bibitem [{\citenamefont {Segovia}\ \emph {et~al.}(2013)\citenamefont
  {Segovia}, \citenamefont {Entem}, \citenamefont {Fernandez},\ and\
  \citenamefont {Hernandez}}]{Segovia:2013wma}%
  \BibitemOpen
  \bibfield  {author} {\bibinfo {author} {\bibfnamefont {J.}~\bibnamefont
  {Segovia}}, \bibinfo {author} {\bibfnamefont {D.~R.}\ \bibnamefont {Entem}},
  \bibinfo {author} {\bibfnamefont {F.}~\bibnamefont {Fernandez}}, \ and\
  \bibinfo {author} {\bibfnamefont {E.}~\bibnamefont {Hernandez}},\ }\href
  {\doibase 10.1142/S0218301313300269} {\bibfield  {journal} {\bibinfo
  {journal} {Int. J. Mod. Phys.}\ }\textbf {\bibinfo {volume} {E22}},\ \bibinfo
  {pages} {1330026} (\bibinfo {year} {2013})},\ \Eprint
  {http://arxiv.org/abs/1309.6926} {arXiv:1309.6926 [hep-ph]} \BibitemShut
  {NoStop}%
\bibitem [{\citenamefont {Vijande}\ \emph {et~al.}(2005)\citenamefont
  {Vijande}, \citenamefont {Fernandez},\ and\ \citenamefont
  {Valcarce}}]{Vijande:2004he}%
  \BibitemOpen
  \bibfield  {author} {\bibinfo {author} {\bibfnamefont {J.}~\bibnamefont
  {Vijande}}, \bibinfo {author} {\bibfnamefont {F.}~\bibnamefont {Fernandez}},
  \ and\ \bibinfo {author} {\bibfnamefont {A.}~\bibnamefont {Valcarce}},\
  }\href {\doibase 10.1088/0954-3899/31/5/017} {\bibfield  {journal} {\bibinfo
  {journal} {J. Phys.}\ }\textbf {\bibinfo {volume} {G31}},\ \bibinfo {pages}
  {481} (\bibinfo {year} {2005})},\ \Eprint
  {http://arxiv.org/abs/hep-ph/0411299} {arXiv:hep-ph/0411299 [hep-ph]}
  \BibitemShut {NoStop}%
\bibitem [{\citenamefont {Valcarce}\ \emph
  {et~al.}(2005{\natexlab{b}})\citenamefont {Valcarce}, \citenamefont
  {Garcilazo},\ and\ \citenamefont {Vijande}}]{Valcarce:2005rr}%
  \BibitemOpen
  \bibfield  {author} {\bibinfo {author} {\bibfnamefont {A.}~\bibnamefont
  {Valcarce}}, \bibinfo {author} {\bibfnamefont {H.}~\bibnamefont {Garcilazo}},
  \ and\ \bibinfo {author} {\bibfnamefont {J.}~\bibnamefont {Vijande}},\ }\href
  {\doibase 10.1103/PhysRevC.72.025206} {\bibfield  {journal} {\bibinfo
  {journal} {Phys. Rev.}\ }\textbf {\bibinfo {volume} {C72}},\ \bibinfo {pages}
  {025206} (\bibinfo {year} {2005}{\natexlab{b}})},\ \Eprint
  {http://arxiv.org/abs/hep-ph/0507297} {arXiv:hep-ph/0507297 [hep-ph]}
  \BibitemShut {NoStop}%
\bibitem [{\citenamefont {Entem}\ and\ \citenamefont
  {Fernandez}(2006)}]{Entem:2006dt}%
  \BibitemOpen
  \bibfield  {author} {\bibinfo {author} {\bibfnamefont {D.~R.}\ \bibnamefont
  {Entem}}\ and\ \bibinfo {author} {\bibfnamefont {F.}~\bibnamefont
  {Fernandez}},\ }\href {\doibase 10.1103/PhysRevC.73.045214} {\bibfield
  {journal} {\bibinfo  {journal} {Phys. Rev.}\ }\textbf {\bibinfo {volume}
  {C73}},\ \bibinfo {pages} {045214} (\bibinfo {year} {2006})}\BibitemShut
  {NoStop}%
\bibitem [{\citenamefont {Valcarce}\ \emph {et~al.}(2008)\citenamefont
  {Valcarce}, \citenamefont {Garcilazo},\ and\ \citenamefont
  {Vijande}}]{Valcarce:2008dr}%
  \BibitemOpen
  \bibfield  {author} {\bibinfo {author} {\bibfnamefont {A.}~\bibnamefont
  {Valcarce}}, \bibinfo {author} {\bibfnamefont {H.}~\bibnamefont {Garcilazo}},
  \ and\ \bibinfo {author} {\bibfnamefont {J.}~\bibnamefont {Vijande}},\ }\href
  {\doibase 10.1140/epja/i2008-10616-4} {\bibfield  {journal} {\bibinfo
  {journal} {Eur. Phys. J.}\ }\textbf {\bibinfo {volume} {A37}},\ \bibinfo
  {pages} {217} (\bibinfo {year} {2008})},\ \Eprint
  {http://arxiv.org/abs/0807.2973} {arXiv:0807.2973 [hep-ph]} \BibitemShut
  {NoStop}%
\bibitem [{\citenamefont {Segovia}\ \emph
  {et~al.}(2008{\natexlab{a}})\citenamefont {Segovia}, \citenamefont {Entem},\
  and\ \citenamefont {Fernandez}}]{Segovia:2008zza}%
  \BibitemOpen
  \bibfield  {author} {\bibinfo {author} {\bibfnamefont {J.}~\bibnamefont
  {Segovia}}, \bibinfo {author} {\bibfnamefont {D.~R.}\ \bibnamefont {Entem}},
  \ and\ \bibinfo {author} {\bibfnamefont {F.}~\bibnamefont {Fernandez}},\
  }\href {\doibase 10.1016/j.physletb.2008.02.051} {\bibfield  {journal}
  {\bibinfo  {journal} {Phys. Lett.}\ }\textbf {\bibinfo {volume} {B662}},\
  \bibinfo {pages} {33} (\bibinfo {year} {2008}{\natexlab{a}})}\BibitemShut
  {NoStop}%
\bibitem [{\citenamefont {Segovia}\ \emph {et~al.}(2009)\citenamefont
  {Segovia}, \citenamefont {Yasser}, \citenamefont {Entem},\ and\ \citenamefont
  {Fernandez}}]{Segovia:2009zz}%
  \BibitemOpen
  \bibfield  {author} {\bibinfo {author} {\bibfnamefont {J.}~\bibnamefont
  {Segovia}}, \bibinfo {author} {\bibfnamefont {A.~M.}\ \bibnamefont {Yasser}},
  \bibinfo {author} {\bibfnamefont {D.~R.}\ \bibnamefont {Entem}}, \ and\
  \bibinfo {author} {\bibfnamefont {F.}~\bibnamefont {Fernandez}},\ }\href
  {\doibase 10.1103/PhysRevD.80.054017} {\bibfield  {journal} {\bibinfo
  {journal} {Phys. Rev.}\ }\textbf {\bibinfo {volume} {D80}},\ \bibinfo {pages}
  {054017} (\bibinfo {year} {2009})}\BibitemShut {NoStop}%
\bibitem [{\citenamefont {Ortega}\ \emph {et~al.}(2011)\citenamefont {Ortega},
  \citenamefont {Entem},\ and\ \citenamefont {Fernandez}}]{Ortega:2011zza}%
  \BibitemOpen
  \bibfield  {author} {\bibinfo {author} {\bibfnamefont {P.~G.}\ \bibnamefont
  {Ortega}}, \bibinfo {author} {\bibfnamefont {D.~R.}\ \bibnamefont {Entem}}, \
  and\ \bibinfo {author} {\bibfnamefont {F.}~\bibnamefont {Fernandez}},\ }\href
  {\doibase 10.1016/j.physletb.2010.12.034} {\bibfield  {journal} {\bibinfo
  {journal} {Phys. Lett.}\ }\textbf {\bibinfo {volume} {B696}},\ \bibinfo
  {pages} {352} (\bibinfo {year} {2011})}\BibitemShut {NoStop}%
\bibitem [{\citenamefont {Segovia}\ \emph {et~al.}(2011)\citenamefont
  {Segovia}, \citenamefont {Entem},\ and\ \citenamefont
  {Fernandez}}]{Segovia:2011zza}%
  \BibitemOpen
  \bibfield  {author} {\bibinfo {author} {\bibfnamefont {J.}~\bibnamefont
  {Segovia}}, \bibinfo {author} {\bibfnamefont {D.~R.}\ \bibnamefont {Entem}},
  \ and\ \bibinfo {author} {\bibfnamefont {F.}~\bibnamefont {Fernandez}},\
  }\bibfield  {booktitle} {\emph {\bibinfo {booktitle} {{Proceedings, 11th
  International Workshop on Meson Production, Properties and Interaction (MESON
  2010): Cracow, Poland, June 10-15, 2010}}},\ }\href {\doibase
  10.1103/PhysRevD.83.114018} {\bibfield  {journal} {\bibinfo  {journal} {Phys.
  Rev.}\ }\textbf {\bibinfo {volume} {D83}},\ \bibinfo {pages} {114018}
  (\bibinfo {year} {2011})}\BibitemShut {NoStop}%
\bibitem [{\citenamefont {Segovia}\ \emph {et~al.}(2015)\citenamefont
  {Segovia}, \citenamefont {Entem},\ and\ \citenamefont
  {Fernandez}}]{Segovia:2015dia}%
  \BibitemOpen
  \bibfield  {author} {\bibinfo {author} {\bibfnamefont {J.}~\bibnamefont
  {Segovia}}, \bibinfo {author} {\bibfnamefont {D.~R.}\ \bibnamefont {Entem}},
  \ and\ \bibinfo {author} {\bibfnamefont {F.}~\bibnamefont {Fernandez}},\
  }\href {\doibase 10.1103/PhysRevD.91.094020} {\bibfield  {journal} {\bibinfo
  {journal} {Phys. Rev.}\ }\textbf {\bibinfo {volume} {D91}},\ \bibinfo {pages}
  {094020} (\bibinfo {year} {2015})},\ \Eprint
  {http://arxiv.org/abs/1502.03827} {arXiv:1502.03827 [hep-ph]} \BibitemShut
  {NoStop}%
\bibitem [{\citenamefont {Segovia}\ \emph {et~al.}(2016)\citenamefont
  {Segovia}, \citenamefont {Ortega}, \citenamefont {Entem},\ and\ \citenamefont
  {Fernández}}]{Segovia:2016xqb}%
  \BibitemOpen
  \bibfield  {author} {\bibinfo {author} {\bibfnamefont {J.}~\bibnamefont
  {Segovia}}, \bibinfo {author} {\bibfnamefont {P.~G.}\ \bibnamefont {Ortega}},
  \bibinfo {author} {\bibfnamefont {D.~R.}\ \bibnamefont {Entem}}, \ and\
  \bibinfo {author} {\bibfnamefont {F.}~\bibnamefont {Fernández}},\ }\href
  {\doibase 10.1103/PhysRevD.93.074027} {\bibfield  {journal} {\bibinfo
  {journal} {Phys. Rev.}\ }\textbf {\bibinfo {volume} {D93}},\ \bibinfo {pages}
  {074027} (\bibinfo {year} {2016})},\ \Eprint
  {http://arxiv.org/abs/1601.05093} {arXiv:1601.05093 [hep-ph]} \BibitemShut
  {NoStop}%
\bibitem [{\citenamefont {Ortega}\ \emph {et~al.}(2016)\citenamefont {Ortega},
  \citenamefont {Segovia}, \citenamefont {Entem},\ and\ \citenamefont
  {Fernandez}}]{Ortega:2016mms}%
  \BibitemOpen
  \bibfield  {author} {\bibinfo {author} {\bibfnamefont {P.~G.}\ \bibnamefont
  {Ortega}}, \bibinfo {author} {\bibfnamefont {J.}~\bibnamefont {Segovia}},
  \bibinfo {author} {\bibfnamefont {D.~R.}\ \bibnamefont {Entem}}, \ and\
  \bibinfo {author} {\bibfnamefont {F.}~\bibnamefont {Fernandez}},\ }\href
  {\doibase 10.1103/PhysRevD.94.074037} {\bibfield  {journal} {\bibinfo
  {journal} {Phys. Rev.}\ }\textbf {\bibinfo {volume} {D94}},\ \bibinfo {pages}
  {074037} (\bibinfo {year} {2016})},\ \Eprint
  {http://arxiv.org/abs/1603.07000} {arXiv:1603.07000 [hep-ph]} \BibitemShut
  {NoStop}%
\bibitem [{\citenamefont {Ortega}\ \emph
  {et~al.}(2017{\natexlab{a}})\citenamefont {Ortega}, \citenamefont {Segovia},
  \citenamefont {Entem},\ and\ \citenamefont {Fernández}}]{Ortega:2016pgg}%
  \BibitemOpen
  \bibfield  {author} {\bibinfo {author} {\bibfnamefont {P.~G.}\ \bibnamefont
  {Ortega}}, \bibinfo {author} {\bibfnamefont {J.}~\bibnamefont {Segovia}},
  \bibinfo {author} {\bibfnamefont {D.~R.}\ \bibnamefont {Entem}}, \ and\
  \bibinfo {author} {\bibfnamefont {F.}~\bibnamefont {Fernández}},\ }\href
  {\doibase 10.1103/PhysRevD.95.034010} {\bibfield  {journal} {\bibinfo
  {journal} {Phys. Rev.}\ }\textbf {\bibinfo {volume} {D95}},\ \bibinfo {pages}
  {034010} (\bibinfo {year} {2017}{\natexlab{a}})},\ \Eprint
  {http://arxiv.org/abs/1612.04826} {arXiv:1612.04826 [hep-ph]} \BibitemShut
  {NoStop}%
\bibitem [{\citenamefont {Ortega}\ \emph {et~al.}(2018)\citenamefont {Ortega},
  \citenamefont {Segovia}, \citenamefont {Entem},\ and\ \citenamefont
  {Fernández}}]{Ortega:2017qmg}%
  \BibitemOpen
  \bibfield  {author} {\bibinfo {author} {\bibfnamefont {P.~G.}\ \bibnamefont
  {Ortega}}, \bibinfo {author} {\bibfnamefont {J.}~\bibnamefont {Segovia}},
  \bibinfo {author} {\bibfnamefont {D.~R.}\ \bibnamefont {Entem}}, \ and\
  \bibinfo {author} {\bibfnamefont {F.}~\bibnamefont {Fernández}},\ }\href
  {\doibase 10.1016/j.physletb.2018.01.005} {\bibfield  {journal} {\bibinfo
  {journal} {Phys. Lett.}\ }\textbf {\bibinfo {volume} {B778}},\ \bibinfo
  {pages} {1} (\bibinfo {year} {2018})},\ \Eprint
  {http://arxiv.org/abs/1706.02639} {arXiv:1706.02639 [hep-ph]} \BibitemShut
  {NoStop}%
\bibitem [{\citenamefont {Ortega}\ \emph
  {et~al.}(2013{\natexlab{a}})\citenamefont {Ortega}, \citenamefont {Entem},\
  and\ \citenamefont {Fernandez}}]{Ortega:2012cx}%
  \BibitemOpen
  \bibfield  {author} {\bibinfo {author} {\bibfnamefont {P.~G.}\ \bibnamefont
  {Ortega}}, \bibinfo {author} {\bibfnamefont {D.~R.}\ \bibnamefont {Entem}}, \
  and\ \bibinfo {author} {\bibfnamefont {F.}~\bibnamefont {Fernandez}},\ }\href
  {\doibase 10.1016/j.physletb.2012.12.025} {\bibfield  {journal} {\bibinfo
  {journal} {Phys. Lett.}\ }\textbf {\bibinfo {volume} {B718}},\ \bibinfo
  {pages} {1381} (\bibinfo {year} {2013}{\natexlab{a}})},\ \Eprint
  {http://arxiv.org/abs/1210.2633} {arXiv:1210.2633 [hep-ph]} \BibitemShut
  {NoStop}%
\bibitem [{\citenamefont {Ortega}\ \emph
  {et~al.}(2014{\natexlab{a}})\citenamefont {Ortega}, \citenamefont {Entem},\
  and\ \citenamefont {Fernández}}]{Ortega:2014eoa}%
  \BibitemOpen
  \bibfield  {author} {\bibinfo {author} {\bibfnamefont {P.~G.}\ \bibnamefont
  {Ortega}}, \bibinfo {author} {\bibfnamefont {D.~R.}\ \bibnamefont {Entem}}, \
  and\ \bibinfo {author} {\bibfnamefont {F.}~\bibnamefont {Fernández}},\
  }\href {\doibase 10.1103/PhysRevD.90.114013} {\bibfield  {journal} {\bibinfo
  {journal} {Phys. Rev.}\ }\textbf {\bibinfo {volume} {D90}},\ \bibinfo {pages}
  {114013} (\bibinfo {year} {2014}{\natexlab{a}})}\BibitemShut {NoStop}%
\bibitem [{\citenamefont {Ortega}\ \emph
  {et~al.}(2014{\natexlab{b}})\citenamefont {Ortega}, \citenamefont {Entem},\
  and\ \citenamefont {Fernández}}]{Ortega:2014fha}%
  \BibitemOpen
  \bibfield  {author} {\bibinfo {author} {\bibfnamefont {P.~G.}\ \bibnamefont
  {Ortega}}, \bibinfo {author} {\bibfnamefont {D.~R.}\ \bibnamefont {Entem}}, \
  and\ \bibinfo {author} {\bibfnamefont {F.}~\bibnamefont {Fernández}},\
  }\href {\doibase 10.1016/j.physletb.2013.12.058} {\bibfield  {journal}
  {\bibinfo  {journal} {Phys. Lett.}\ }\textbf {\bibinfo {volume} {B729}},\
  \bibinfo {pages} {24} (\bibinfo {year} {2014}{\natexlab{b}})}\BibitemShut
  {NoStop}%
\bibitem [{\citenamefont {Ortega}\ \emph
  {et~al.}(2017{\natexlab{b}})\citenamefont {Ortega}, \citenamefont {Entem},\
  and\ \citenamefont {Fernández}}]{Ortega:2016syt}%
  \BibitemOpen
  \bibfield  {author} {\bibinfo {author} {\bibfnamefont {P.~G.}\ \bibnamefont
  {Ortega}}, \bibinfo {author} {\bibfnamefont {D.~R.}\ \bibnamefont {Entem}}, \
  and\ \bibinfo {author} {\bibfnamefont {F.}~\bibnamefont {Fernández}},\
  }\href {\doibase 10.1016/j.physletb.2016.11.008} {\bibfield  {journal}
  {\bibinfo  {journal} {Phys. Lett.}\ }\textbf {\bibinfo {volume} {B764}},\
  \bibinfo {pages} {207} (\bibinfo {year} {2017}{\natexlab{b}})},\ \Eprint
  {http://arxiv.org/abs/1606.06148} {arXiv:1606.06148 [hep-ph]} \BibitemShut
  {NoStop}%
\bibitem [{\citenamefont {Ortega}\ \emph {et~al.}(2010)\citenamefont {Ortega},
  \citenamefont {Segovia}, \citenamefont {Entem},\ and\ \citenamefont
  {Fernandez}}]{Ortega:2009hj}%
  \BibitemOpen
  \bibfield  {author} {\bibinfo {author} {\bibfnamefont {P.~G.}\ \bibnamefont
  {Ortega}}, \bibinfo {author} {\bibfnamefont {J.}~\bibnamefont {Segovia}},
  \bibinfo {author} {\bibfnamefont {D.~R.}\ \bibnamefont {Entem}}, \ and\
  \bibinfo {author} {\bibfnamefont {F.}~\bibnamefont {Fernandez}},\ }\href
  {\doibase 10.1103/PhysRevD.81.054023} {\bibfield  {journal} {\bibinfo
  {journal} {Phys. Rev.}\ }\textbf {\bibinfo {volume} {D81}},\ \bibinfo {pages}
  {054023} (\bibinfo {year} {2010})},\ \Eprint {http://arxiv.org/abs/0907.3997}
  {arXiv:0907.3997 [hep-ph]} \BibitemShut {NoStop}%
\bibitem [{\citenamefont {Ortega}\ \emph
  {et~al.}(2013{\natexlab{b}})\citenamefont {Ortega}, \citenamefont {Entem},\
  and\ \citenamefont {Fernandez}}]{Ortega:2012rs}%
  \BibitemOpen
  \bibfield  {author} {\bibinfo {author} {\bibfnamefont {P.~G.}\ \bibnamefont
  {Ortega}}, \bibinfo {author} {\bibfnamefont {D.~R.}\ \bibnamefont {Entem}}, \
  and\ \bibinfo {author} {\bibfnamefont {F.}~\bibnamefont {Fernandez}},\ }\href
  {\doibase 10.1088/0954-3899/40/6/065107} {\bibfield  {journal} {\bibinfo
  {journal} {J. Phys.}\ }\textbf {\bibinfo {volume} {G40}},\ \bibinfo {pages}
  {065107} (\bibinfo {year} {2013}{\natexlab{b}})},\ \Eprint
  {http://arxiv.org/abs/1205.1699} {arXiv:1205.1699 [hep-ph]} \BibitemShut
  {NoStop}%
\bibitem [{\citenamefont {Diakonov}(2003)}]{Diakonov:2002fq}%
  \BibitemOpen
  \bibfield  {author} {\bibinfo {author} {\bibfnamefont {D.}~\bibnamefont
  {Diakonov}},\ }\href {\doibase 10.1016/S0146-6410(03)90014-7} {\bibfield
  {journal} {\bibinfo  {journal} {Prog. Part. Nucl. Phys.}\ }\textbf {\bibinfo
  {volume} {51}},\ \bibinfo {pages} {173} (\bibinfo {year} {2003})},\ \Eprint
  {http://arxiv.org/abs/hep-ph/0212026} {arXiv:hep-ph/0212026 [hep-ph]}
  \BibitemShut {NoStop}%
\bibitem [{\citenamefont {Bali}\ \emph {et~al.}(2005)\citenamefont {Bali},
  \citenamefont {Neff}, \citenamefont {Duessel}, \citenamefont {Lippert},\ and\
  \citenamefont {Schilling}}]{Bali:2005fu}%
  \BibitemOpen
  \bibfield  {author} {\bibinfo {author} {\bibfnamefont {G.~S.}\ \bibnamefont
  {Bali}}, \bibinfo {author} {\bibfnamefont {H.}~\bibnamefont {Neff}}, \bibinfo
  {author} {\bibfnamefont {T.}~\bibnamefont {Duessel}}, \bibinfo {author}
  {\bibfnamefont {T.}~\bibnamefont {Lippert}}, \ and\ \bibinfo {author}
  {\bibfnamefont {K.}~\bibnamefont {Schilling}} (\bibinfo {collaboration}
  {SESAM}),\ }\href {\doibase 10.1103/PhysRevD.71.114513} {\bibfield  {journal}
  {\bibinfo  {journal} {Phys. Rev.}\ }\textbf {\bibinfo {volume} {D71}},\
  \bibinfo {pages} {114513} (\bibinfo {year} {2005})},\ \Eprint
  {http://arxiv.org/abs/hep-lat/0505012} {arXiv:hep-lat/0505012 [hep-lat]}
  \BibitemShut {NoStop}%
\bibitem [{\citenamefont {Segovia}\ \emph
  {et~al.}(2008{\natexlab{b}})\citenamefont {Segovia}, \citenamefont {Yasser},
  \citenamefont {Entem},\ and\ \citenamefont {Fernandez}}]{Segovia:2008zz}%
  \BibitemOpen
  \bibfield  {author} {\bibinfo {author} {\bibfnamefont {J.}~\bibnamefont
  {Segovia}}, \bibinfo {author} {\bibfnamefont {A.~M.}\ \bibnamefont {Yasser}},
  \bibinfo {author} {\bibfnamefont {D.~R.}\ \bibnamefont {Entem}}, \ and\
  \bibinfo {author} {\bibfnamefont {F.}~\bibnamefont {Fernandez}},\ }\href
  {\doibase 10.1103/PhysRevD.78.114033} {\bibfield  {journal} {\bibinfo
  {journal} {Phys. Rev.}\ }\textbf {\bibinfo {volume} {D78}},\ \bibinfo {pages}
  {114033} (\bibinfo {year} {2008}{\natexlab{b}})}\BibitemShut {NoStop}%
\bibitem [{\citenamefont {Wheeler}(1937)}]{Wheeler:1937zza}%
  \BibitemOpen
  \bibfield  {author} {\bibinfo {author} {\bibfnamefont {J.~A.}\ \bibnamefont
  {Wheeler}},\ }\href {\doibase 10.1103/PhysRev.52.1083} {\bibfield  {journal}
  {\bibinfo  {journal} {Phys. Rev.}\ }\textbf {\bibinfo {volume} {52}},\
  \bibinfo {pages} {1083} (\bibinfo {year} {1937})}\BibitemShut {NoStop}%
\bibitem [{\citenamefont {Hiyama}\ \emph {et~al.}(2003)\citenamefont {Hiyama},
  \citenamefont {Kino},\ and\ \citenamefont {Kamimura}}]{Hiyama:2003cu}%
  \BibitemOpen
  \bibfield  {author} {\bibinfo {author} {\bibfnamefont {E.}~\bibnamefont
  {Hiyama}}, \bibinfo {author} {\bibfnamefont {Y.}~\bibnamefont {Kino}}, \ and\
  \bibinfo {author} {\bibfnamefont {M.}~\bibnamefont {Kamimura}},\ }\href
  {\doibase 10.1016/S0146-6410(03)90015-9} {\bibfield  {journal} {\bibinfo
  {journal} {Prog. Part. Nucl. Phys.}\ }\textbf {\bibinfo {volume} {51}},\
  \bibinfo {pages} {223} (\bibinfo {year} {2003})}\BibitemShut {NoStop}%
\bibitem [{\citenamefont {Machleidt}(1993)}]{Machleidt:1003bo}%
  \BibitemOpen
  \bibfield  {author} {\bibinfo {author} {\bibfnamefont {R.}~\bibnamefont
  {Machleidt}},\ }\href@noop {} {\emph {\bibinfo {title} {Computational Nuclear
  Physics 2: Nuclear Reactions}}},\ \bibinfo {edition} {edited by k. langanke,
  j. a. maruhn, and s. e. koonin}\ ed.\ (\bibinfo  {publisher}
  {Springer-Verlag},\ \bibinfo {address} {Berlin},\ \bibinfo {year} {1993})\
  pp.\ \bibinfo {pages} {1--29}\BibitemShut {NoStop}%
\bibitem [{\citenamefont {Yuan}(2018)}]{Yuan:2018inv}%
  \BibitemOpen
  \bibfield  {author} {\bibinfo {author} {\bibfnamefont {C.-Z.}\ \bibnamefont
  {Yuan}},\ }\href {\doibase 10.1142/S0217751X18300181} {\bibfield  {journal}
  {\bibinfo  {journal} {Int. J. Mod. Phys.}\ }\textbf {\bibinfo {volume}
  {A33}},\ \bibinfo {pages} {1830018} (\bibinfo {year} {2018})},\ \Eprint
  {http://arxiv.org/abs/1808.01570} {arXiv:1808.01570 [hep-ex]} \BibitemShut
  {NoStop}%
\bibitem [{\citenamefont {He}(2015)}]{He:2015mja}%
  \BibitemOpen
  \bibfield  {author} {\bibinfo {author} {\bibfnamefont {J.}~\bibnamefont
  {He}},\ }\href {\doibase 10.1103/PhysRevD.92.034004} {\bibfield  {journal}
  {\bibinfo  {journal} {Phys. Rev.}\ }\textbf {\bibinfo {volume} {D92}},\
  \bibinfo {pages} {034004} (\bibinfo {year} {2015})},\ \Eprint
  {http://arxiv.org/abs/1505.05379} {arXiv:1505.05379 [hep-ph]} \BibitemShut
  {NoStop}%
\end{thebibliography}%

\end{document}